\documentclass[
    a4paper,
    reprint,
    superscriptaddress,
    amsmath,amssymb,
    aip,
]{revtex4-2}

\usepackage{my-aip}
\usepackage{xr}
\IfFileExists{supplement.aux}{\externaldocument{supplement}}{%
  \IfFileExists{supplement-xref.tex}{\input{supplement-xref.tex}}{}}
\graphicspath{{figures/}}

\begin{document}

\title{
    Nonequilibrium quantum states emerging from flippable spins
    in random networks
}

\author{Yoshiaki Horiike\,\orcidlink{0009-0000-2010-2598}}
\email{yoshi.h@nagoya-u.jp}
\affiliation{
    Department of Applied Physics,
    Nagoya University,
    Nagoya, Japan
}

\author{Yuki Kawaguchi\,\orcidlink{0000-0003-1668-6484}}
\affiliation{
    Department of Applied Physics,
    Nagoya University,
    Nagoya, Japan
}
\affiliation{
    Research Center for Crystalline Materials Engineering,
    Nagoya University,
    Nagoya, Japan
}

\date{17 October 2026}

\begin{abstract}
    Recent advances in programmable quantum
    simulators~\cite{Blatt2012,Browaeys2020,Altman2021,Daley2022}
    have enabled the realization of irregular interaction networks,
    which natural crystal lattices cannot provide,
    together with time-resolved observation of quantum many-body
    dynamics~\cite{Kinoshita2006,Bernien2017,Bluvstein2021,King2022,Zhang2023,Yao2023,Adler2024}.
    Understanding the nonequilibrium states that emerge from such networks
    has only just
    begun~\cite{Tindall2022,Searle2024,Grabarits2025,Hashizume2025},
    and whether quantum many-body scar states can emerge in them remains
    open.
    Here, focusing on a quantum Ising spin system
    with random interaction networks under a weak transverse field,
    we introduce the flippability, the fraction of flippable spins,
    and present phenomenological laws that predict the
    emergence of quantum many-body scar states from the network structure.
    As the flippability increases,
    the Hilbert space becomes less fragmented.
    The largest kinetically accessible subspace shows nonintegrable
    level statistics,
    yet hosts a few eigenstates with abnormally low participation entropy,
    signalling scar states.
    Real-time dynamics from Fock states confirm these signatures:
    systems with higher flippability tend toward thermalization,
    whereas a few initial states show anomalously nonthermal dynamics.
    The flippability thus links the interaction-network structure of the
    Hamiltonian to nonequilibrium many-body dynamics without translational
    invariance,
    and may apply to other kinetically constrained quantum systems.
\end{abstract}

\maketitle

Modern quantum many-body physics fundamentally relies on translational symmetry.
Beginning with the celebrated Bloch theorem and Wannier functions,
the regular structure of the crystal lattice has provided solutions to the
Schr\"odinger equation.
The periodicity of the wave function plays a crucial role in
understanding electrons in solids,
and representing the wave function in wave vector space
forms the foundation of band theory in condensed matter physics.

Recent advances in atomic, molecular, and optical physics have
enabled the creation of isolated artificial quantum many-body systems
without translational symmetry. Several controllable, tunable, or
programmable platforms have been developed for the realization of quantum
computers~\cite{CommitteeOnTechnicalAssessmentOfTheFeasibilityAndImplicationsOfQuantumComputing2019},
particularly quantum simulators~\cite{Altman2021,Daley2022}.
These platforms enable the implementation of arbitrary interactions among
their constituent elements, which are difficult to realize in natural crystals.
Examples of such platforms include trapped
ions~\cite{Porras2004,Kim2010,Britton2012,Morong2021},
Rydberg
atoms~\cite{Browaeys2020,Labuhn2016,Bluvstein2021},
atoms in cavities~\cite{Hung2016,Periwal2021}, nuclear
spins~\cite{Randall2021}, superconducting
qubits~\cite{Johnson2011,King2022,King2023,King2025,Zhang2023,Yao2023},
and parametric
oscillators~\cite{Inagaki2016,McMahon2016,Honjo2021},
to list a few.
These emerging artificial quantum many-body systems are opening new
avenues in combinatorial optimization~\cite{Mohseni2022}, quantum
gravity~\cite{Kollar2019,Periwal2021}, and,
obviously, quantum many-body
theory~\cite{Tindall2022,Searle2024,Grabarits2025,Defenu2023,Defenu2024,Hashizume2025}.

Such programmable quantum many-body systems enable detailed
observation of microscopic behaviour.
Notably, quantum simulators are opening new avenues for investigating
time-dependent
behaviour~\cite{Kinoshita2006,Bernien2017,Turner2018,Turner2018a,Bluvstein2021,Morong2021,Randall2021,King2022,Zhang2023,Yao2023,Adler2024}
and nonequilibrium
phenomena~\cite{Polkovnikov2011,Abanin2019}
that are intractable with conventional experiments with
crystalline solids.
We can now directly observe and test theoretical predictions
regarding quantum many-body nonequilibrium dynamics.

Given these recent experimental advances, a new question arises:
How does the structure of the interaction network determine quantum
many-body nonequilibrium phenomena?
Previous studies have shown that network structure influences quantum
many-body behaviours such as
quantum many-body scar states~\cite{Bluvstein2021,Jeevanesan2023,Surace2023},
quantum phase transitions~\cite{Tindall2022,Searle2024,Hashizume2025}
(see, however, ref.~\cite{Ostilli2020}),
and chaotic behaviour~\cite{Grabarits2025}.
Additionally, long-range interaction networks~\cite{Defenu2023,Defenu2024}
have attracted interest in both equilibrium and nonequilibrium quantum
many-body phenomena.
Despite these advances, our understanding of how network structure
relates to quantum many-body phenomena remains limited to either
density of random network
structures~\cite{Tindall2022,Searle2024,Grabarits2025,Hashizume2025} or
extensions of regular lattices~\cite{Defenu2023,Defenu2024}:
the former reveals the average behaviour across various network structures,
while the latter relies on the periodicity of the interaction network.
Consequently, it remains challenging to predict the quantum many-body
phenomena,
such as quantum many-body scar states and quantum many-body localization,
based on a given network with arbitrary structure.
Is there any simple principle of network structure that governs the
quantum many-body nonequilibrium phenomena?

Here, by focusing on the quantum many-body scar state of
an Ising spin system under a weak transverse field,
we reveal phenomenological laws that
predict the existence of the quantum many-body nonequilibrium states
from the given network structure.
We introduce the concept of flippability,
which quantifies the fraction of flippable spins of a given Hamiltonian.
With numerical analysis of the ensemble of Erd\H{o}s--R\'enyi random
networks (with fixed edge probability) and random longitudinal fields,
we show that flippability controls the fragmentation of the Hilbert space:
as the flippability increases, the size of the largest
kinetically accessible subspace increases.
We investigate the largest subspace and find it in the nonintegrable
regime, yet it hosts eigenstates with nontrivially low participation
entropy, indicating the existence of states violating the eigenstate
thermalization hypothesis.
Time-evolution simulations initialized from Fock states further
reveal that the fraction of quantum many-body scar states within the
largest kinetically accessible subspace increases monotonically with
flippability in the thermalizing regime.
Extrapolations to the thermodynamic limit suggest that the critical
flippability separating recurrent and thermalizing dynamics approaches zero,
while the fraction of scar states within the largest kinetically
accessible subspace remains finite. These findings identify
flippability as a predictor of both nonequilibrium dynamics and
many-body scarring in quantum Ising models on random interaction networks.

\begin{figure*}[tb]
    \centering
    \includegraphics{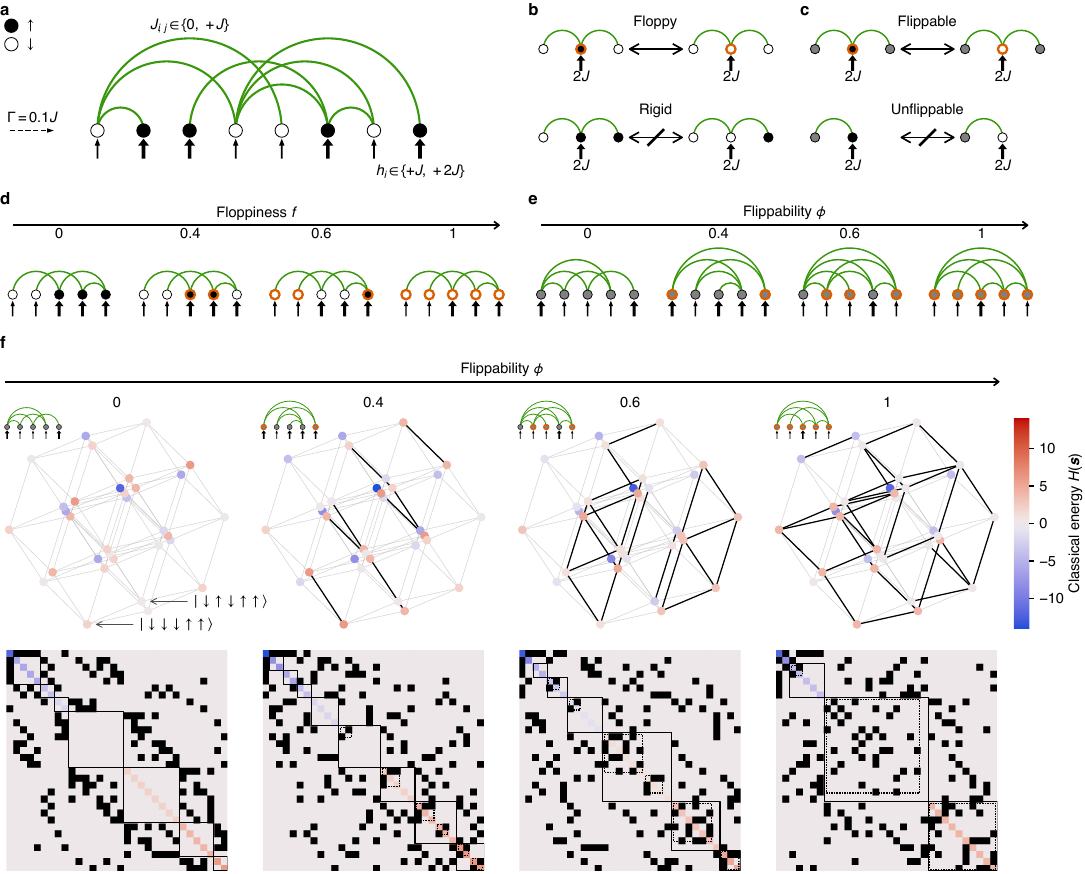}
    \bcaption{
        Model, floppiness and flippability.
    }{
        \textbf{a},
        An example ferromagnetic Erd\H{o}s--R\'enyi network with
        random longitudinal fields
        [equation~\eqref{eq:Hamiltonian}].
        Filled (open) circles denote up (down) spins;
        edges indicate interactions $J_{i,j}$;
        arrow widths indicate the longitudinal fields $|h_i|$;
        the dashed arrow is the transverse field.
        \textbf{b},
        Top: the central spin is floppy because its net local
        longitudinal field vanishes [equation~\eqref{eq:local-field}],
        so it can flip without changing the energy.
        Bottom: a rigid spin, whose net local field is nonzero.
        \textbf{c},
        Top: a flippable spin, whose local field
        vanishes when both interacting spins point down.
        Bottom: with the same $h_i = 2 J$ but one interacting neighbour,
        the local field cannot vanish, so the spin is unflippable.
        \textbf{d},
        Examples of floppiness [equation~\eqref{eq:floppiness}] for one network
        across spin configurations.
        \textbf{e},
        Examples of flippability [equation~\eqref{eq:flippability}] for four
        network--field realizations.
        The interacting spins are shown in grey because flippability
        is independent of their orientations.
        Floppy ($g_k = 0$) and flippable ($\gamma_k = 0$) spins
        are outlined in orange.
        \textbf{f},
        Fock-space (hypercubic) landscapes of the networks of
        \textbf{e} (top left of each landscape).
        Top: all $2^N$ Fock states projected to two dimensions
        and coloured by classical energy;
        edges connect states one spin flip apart,
        thick black for equal-energy (kinetically permitted) hops
        and thin grey for off-resonant ones.
        Bottom: the Hamiltonians [equation~\eqref{eq:Hamiltonian}] in the
        Fock basis, ordered by classical energy;
        solid boxes mark equal-energy sectors
        and dashed boxes the kinetically accessible subspaces within them.
        Diagonal entries share the vertex colour scale;
        nonzero off-diagonal (transverse-field) entries are black.
    }
    \label{fig:fig-01}
\end{figure*}

\section*{Model, floppiness and flippability}
We consider the transverse-field Ising model,
which can be realized in various platforms of quantum
simulators~\cite{Periwal2021,King2023,King2025}.
The Hamiltonian of the $N$-spin quantum Ising model is given by
\begin{equation}
    \hat{\mathcal{H}}
    \coloneqq
    -
    \frac{1}{2}
    \sum_{i=1}^N
    \sum_{j=1}^N
    J_{i,j} \hat{Z}_i \hat{Z}_j
    -
    \sum_{i=1}^N h_i \hat{Z}_i
    -
    \varGamma
    \sum_{i=1}^N  \hat{X}_i
    \label{eq:Hamiltonian}
    ,
\end{equation}
where
$
\hat{Z}_i
$
and
$
\hat{X}_i
$
are the Pauli matrices acting on the $i$th spin.
The coupling strength between spins $i$ and $j$ is specified by
$J_{i,j}=J_{j,i}$, the element of the interaction matrix
$\bm{J} \in \{0, +J\}^{N\times N}$,
with constant $J \in \mathbb{R}_{>0}$
and no self-interaction (i.e., $J_{i,i}=0$ for all $i$)
(\nameref{sec:random-network-generation} in \nameref{sec:methods}).
The non-interacting terms are given by $h_i \in \{+J, +2J\}$,
which are the longitudinal fields acting on the $i$th spin,
and the uniform and weak transverse field is set to be $\varGamma = 0.1J$.
We consider the transverse field as a perturbation to the
classical Ising model.
For the network generation,
we employ random Erd\H{o}s--R\'enyi networks with edge probability
$p_\mathrm{ER} = 0.5$ and $N \in \{10, 11, 12, 13, 14, 15\}$ spins,
but we exclude realizations in which one or more spins are disconnected from
the rest of the system.
The longitudinal fields are drawn randomly from $h_i \in \{+J, +2J\}$.
These random longitudinal fields can be regarded as interactions from
frozen spins outside the system~\cite{Yoshinaga2022}.
See Fig.~\ref{fig:fig-01}a for the model setup with an example of the
interaction network and the longitudinal fields.
Note that we also consider the $\pm J$ model,
where the interaction network is given by
$\bm{J} \in \{-J, 0, +J\}^{N\times N}$,
in the Supplementary Information,
and we confirm that the results are qualitatively the same as those of the
ferromagnetic model above.

The Hamiltonian in equation~\eqref{eq:Hamiltonian} can be interpreted
as a single quantum particle hopping on the $N$-dimensional hypercube.
To see this, we expand the Hamiltonian in equation~\eqref{eq:Hamiltonian}
in the Fock basis as follows:
\begin{align}
    \hat{\mathcal{H}}
    &=
    \Bigg(
        \sum_{\bm{s}^\prime}
        \ket{\bm{s}^\prime} \! \bra{\bm{s}^\prime}
    \Bigg)
    \hat{\mathcal{H}}
    \Bigg(
        \sum_{\bm{s}}
        \ket{\bm{s}} \! \bra{\bm{s}}
    \Bigg)
    \nonumber
    \\
    &=
    \sum_{\bm{s}}
    H(\bm{s})
    \ket{\bm{s}} \! \bra{\bm{s}}
    +
    \sum_{\bm{s}}
    \sum_{\bm{s}^\prime \neq \bm{s}}
    K(\bm{s}^\prime, \bm{s})
    \ket{\bm{s}^\prime} \! \bra{\bm{s}}
    \nonumber
    \\
    &=
    \sum_{\bm{s}}
    H(\bm{s})
    \ket{\bm{s}} \! \bra{\bm{s}}
    -
    \varGamma
    \sum_{\bm{s}}
    \sum_{k=1}^N
    \ket{\bm{F}_{k} \bm{s}} \! \bra{\bm{s}}
    \label{eq:Anderson-Hamiltonian}
    ,
\end{align}
where
$
\bm{s}
\coloneqq
[
    s_1 \; \cdots \; s_N
]^\top
\in
\{
    \bullet=\;\uparrow\;\coloneqq +1,
    \circ=\;\downarrow\;\coloneqq -1
\}^N
$
is the spin configuration that indexes the Fock state,
and
$
\ket{\bm{s}}
\coloneqq
\bigotimes_{i=1}^N
\ket{s_i}
$
is the corresponding Fock state (or computational-basis state).
By writing
$
\hat{Z}_i
=
\ket{\uparrow_i} \!
\bra{\uparrow_i}
-
\ket{\downarrow_i} \!
\bra{\downarrow_i}
$
and
$
\hat{X}_i
=
\ket{\uparrow_i} \!
\bra{\downarrow_i}
+
\ket{\downarrow_i} \!
\bra{\uparrow_i}
$,
the zero-transverse-field eigenenergy (classical energy)
of a spin configuration is given by
$
H(\bm{s})
\coloneqq
\braket{\bm{s} | \hat{\mathcal{H}} | \bm{s}}
=
-
(1/2)
\sum_{i=1}^N
\sum_{j=1}^N
s_i J_{i,j} s_j
-
\sum_{i=1}^N
s_i h_i
=
-
(1/2)
\bm{s}^\top
\bm{J}
\bm{s}
-
\bm{s}^\top
\bm{h}
$,
where
$
\bm{h}
\coloneqq
[
    h_1 \; \cdots \; h_N
]^\top
\in
\{+J, +2J\}^N
$
is the longitudinal field vector.
The parameter
$
K (\bm{s}^\prime, \bm{s})
\coloneqq
\braket{\bm{s}^\prime | \hat{\mathcal{H}} | \bm{s}}
=
-\varGamma
\sum_{k=1}^{N}
\delta_{\bm{F}_{k} \bm{s}, \bm{s}^\prime}
$
represents the hopping energy
between the Fock states $\ket{\bm{s}}$ and $\ket{\bm{s}^\prime}$ and
allows only single spin flips.
The spin-flip matrix acts on the spin configuration as
$
\bm{F}_{k}
\coloneqq
\bm{I}
-
2
\bm{e}_k \bm{e}_k^\top
$,
which flips the $k$th spin of configuration $\bm{s}$,
$
\bm{F}_{k}
\bm{s}
=
[
    s_1 \; \cdots \; -s_k \; \cdots \; s_N
]^\top
$.
Here, $\bm{e}_k$ is the $k$th standard unit vector in $\mathbb{R}^N$, and
$\bm{I}$ is the identity matrix.
Note that the first term of equation~\eqref{eq:Anderson-Hamiltonian}
depends on the interaction network, whereas the second term does not.
In this representation, the
Hamiltonian~\eqref{eq:Anderson-Hamiltonian} can be interpreted as
describing a single quantum particle hopping on an $N$-dimensional hypercube,
with on-site energies given by $H(\bm{s})$ and hopping amplitude $-\varGamma$.
Thus, equation~\eqref{eq:Anderson-Hamiltonian} is the Hamiltonian of the
Anderson model
of localization~\cite{Anderson1958} or
the Fock-space mapping of many-body systems~\cite{Welsh2018,Yao2023,Roy2025}.
This Anderson-model perspective on Hilbert space has also been
highlighted in the context of
the Grover search algorithm~\cite{Grover1997,Childs2004},
quantum annealing~\cite{Santoro2006,Altshuler2010},
time crystals~\cite{Estarellas2020},
and
many-body localization~\cite{Altshuler2010,Alet2018,Nokkala2024,Roy2025}.
By identifying each Fock state with a vertex of the hypercube
and each transverse field term with a hypercubic edge,
the Hamiltonian in equation~\eqref{eq:Anderson-Hamiltonian}
corresponds to a particle hopping on an $N$-dimensional hypercube.

The key to understanding the hypercubic structure of the Hamiltonian of
equation~\eqref{eq:Anderson-Hamiltonian}
is the floppiness and flippability of the spins.
Under the weak transverse field,
the kinetic constraints arise from the interaction network structure
and the longitudinal fields,
and these constraints are the origin of the Hilbert space fragmentation,
which has been implicitly introduced in the
literature~\cite{King2016,Turner2018,Turner2018a,VanVoorden2020,Yoshinaga2022,Balducci2022,Hart2022}.
A state of spin $k$ is called floppy if $H(\bm{F}_k\bm{s}) = H(\bm{s})$;
otherwise, it is called rigid (Fig.~\ref{fig:fig-01}b).
The fraction of floppy spins defines the floppiness $f$
(Fig.~\ref{fig:fig-01}d), which is
\begin{align}
    f
    &\coloneqq
    \frac{1}{N}
    \sum_{k=1}^N
    \delta_{0, g_k}
    \label{eq:floppiness}
    \\
    g_k
    &\coloneqq
    \sum_{j=1}^N J_{k,j} s_j + h_k
    \label{eq:local-field}
    ,
\end{align}
where $\delta_{i, j}$ is the Kronecker delta and
$g_k$ is the incoming field to spin $k$.
A spin $k$ is called flippable if its local field $g_k$ can vanish
for some configuration of its interacting spins (Fig.~\ref{fig:fig-01}c).
The fraction of flippable spins defines the flippability $\phi$
(Fig.~\ref{fig:fig-01}e), which is defined as
\begin{align}
    \phi
    &\coloneqq
    \frac{1}{N}
    \sum_{k=1}^N
    \delta_{0, \gamma_k}
    \label{eq:flippability}
    \\
    \gamma_k
    &\coloneqq
    \frac{1}{J}
    \Bigg(
        \sum_{j=1}^N |J_{k,j}| + |h_k|
    \Bigg)
    \mathrm{mod}~2
    \label{eq:num-effective-interaction}
    ,
\end{align}
where $\gamma_k$ quantifies whether spin $k$ has an even number of
interactions or not, i.e.,
it counts the number of nonzero interactions by summing the absolute
values of the interaction strengths (adjacency matrix),
$|J_{k,j}| \in \{0, +J\}$,
and the absolute value of the longitudinal field, $|h_k| \in \{+J, +2J\}$.
This definition of flippability applies to models in which the
interaction and longitudinal-field strengths are integer multiples of $J$.
Note that the floppiness $f$ is determined by
the interaction network $\bm{J}$,
the longitudinal fields $\bm{h}$,
and the spin configuration $\bm{s}$,
while the flippability $\phi$ depends only on the parameters of the Hamiltonian,
$\bm{J}$ and $\bm{h}$, and not on the spin configuration $\bm{s}$.
Thus, the flippability $\phi$ characterizes the Hamiltonian.

When the transverse field is weak compared with the interaction and
longitudinal-field energy scales,
the Hamiltonian of equation~\eqref{eq:Hamiltonian} can be approximated by
retaining only resonant spin-flip processes associated with floppy
spins~\cite{Turner2018,Turner2018a,VanVoorden2020,Yoshinaga2022,Peng2022a}.
Specifically, when $|\varGamma| \ll |J|$, spin flips that
change the unperturbed (classical) energy by an amount of order $J$ are
off-resonant and contribute only at higher orders in $\varGamma/J$.
Neglecting these off-resonant processes,
we retain only the matrix elements of the transverse-field term
within each unperturbed-energy
sector~\cite{Turner2018,Turner2018a,VanVoorden2020,Yoshinaga2022,Peng2022a}.
In the Fock basis, the resulting effective Hamiltonian is given by
\begin{align}
    \hat{\mathcal{H}}
    &\approx
    \sum_{\bm{s}}
    H(\bm{s})
    \ket{\bm{s}} \! \bra{\bm{s}}
    -
    \varGamma
    \sum_{\bm{s}}
    \sum_{k=1}^N
    \delta_{H(\bm{s}), H(\bm{F}_{k} \bm{s})}
    \ket{\bm{F}_{k} \bm{s}} \! \bra{\bm{s}}
    \label{eq:approximated-Hamiltonian}
    .
\end{align}
See the Supplementary Information for the derivations of this equienergy
approximation
and Extended Data Fig.~\ref{fig:fig-s01} for its numerical validation.

Thus, in the weak-transverse-field limit, the effective Hamiltonian
allows hopping only between Fock states with equal unperturbed energies,
$H(\bm{F}_k \bm{s}) = H(\bm{s})$, corresponding to spin flips with $g_k = 0$.
The effective Hamiltonian can thus be viewed as describing a quantum
particle hopping on an equienergy surface of the hypercubic energy
landscape~\cite{Farhan2013}, with hopping amplitude $-\varGamma$.
Following ref.~\cite{Roy2025},
we refer to this structure as the equienergy Fock-space landscape.
In the Fock-basis representation, this approximation amounts to
retaining only matrix elements within each equienergy sector,
resulting in a block-diagonal Hamiltonian (Fig.~\ref{fig:fig-01}f).

The equienergy sector is further divided into the kinetically accessible
subspaces.
As shown in Fig.~\ref{fig:fig-01}f, the flippability,
rather than the classical energy value itself,
is the key to understanding the actual hopping of the quantum particle in
the Fock-space landscape,
i.e., the state transition is not allowed even if the unperturbed
energy eigenvalues are the same.
We emphasize that the hopping is only allowed if the two Fock states
are connected by a \emph{single} spin flip of a floppy spin.
Thus, an equienergy sector is further fragmented into kinetically
accessible subspaces,
where the actual state transition happens,
and inter-subspace hopping is forbidden.
In the matrix representation of the Hamiltonian in the Fock basis,
this subspace fragmentation corresponds to the further block-diagonalization of
the equienergy block (Fig.~\ref{fig:fig-01}f).
Although the flippability is not directly equivalent to the
floppiness of the Fock state,
we show, in the rest of this paper,
that the flippability enables the prediction of the size of
the largest kinetically accessible subspace.
In this work, we limit our analysis to the largest equienergy sector
and the largest kinetically accessible subspace within it,
which are the most representative of the Hamiltonian and can be
compared across different realizations of the Hamiltonian.

It is, in general, not trivial to predict the structure of the
subspace-fragmented Hamiltonian from the given interaction network
$\bm{J}$ and longitudinal fields $\bm{h}$.
Our flippability, however, provides simple and effective
phenomenological laws to predict the behaviour of the Hamiltonian of
equation~\eqref{eq:Hamiltonian}:
(1) As the flippability $\phi$ increases, the size of the largest
kinetically accessible subspace also increases;
(2) in the thermodynamic limit, all Hamiltonians become nonintegrable
regardless of the flippability;
(3) nevertheless, the Fock-state-initialized dynamics can exhibit
nonthermal behaviour,
such as quantum many-body scar states,
for Hamiltonians with high flippability.
Our numerical results below support these phenomenological laws.

\begin{figure*}[tb]
    \centering
    \includegraphics{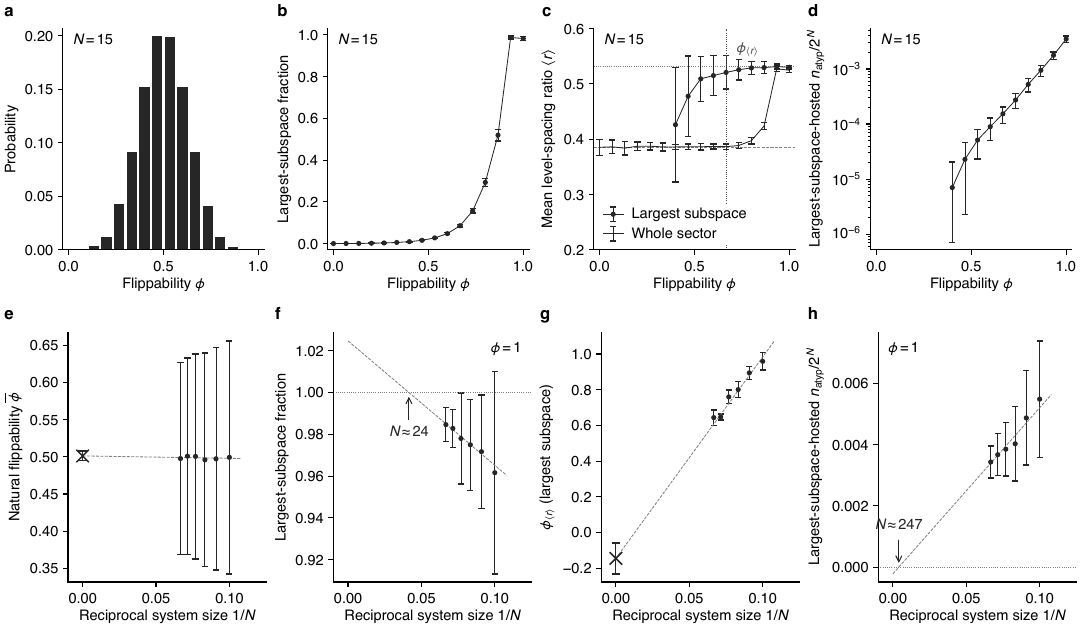}
    \bcaption{
        Static features of the Hamiltonian along flippability and
        their thermodynamic-limit extrapolations.
    }{
        \textbf{a},
        Distribution of flippability $\phi$
        in the Erd\H{o}s--R\'enyi ensemble with random longitudinal fields
        $h_i \in \{+J, +2J\}$ ($N = 15$; $10\,000$ draws).
        Panels \textbf{b}--\textbf{d} and \textbf{f}--\textbf{h}
        instead use equal numbers of Hamiltonians per flippability level
        (\nameref{sec:flippability-balanced-protocol} in~\nameref{sec:methods}).
        \textbf{b},
        Fraction of the largest equienergy sector occupied by its largest
        kinetically accessible subspace versus $\phi$ ($N = 15$).
        \textbf{c},
        Mean level-spacing ratio $\langle r \rangle$ versus
        $\phi$ ($N = 15$) for the whole sector (line) and the largest
        subspace (filled circles; shown where at least half of the
        networks have a largest subspace of at least $20$ levels).
        Dashed and dotted horizontal lines: Poisson ($\approx 0.386$) and
        Gaussian orthogonal ensemble (GOE; $\approx 0.531$)
        limits~\cite{Atas2013};
        dotted vertical line: GOE-confident flippability
        $\phi_{\langle r \rangle}$
        (\nameref{sec:energy-level-statistics} in \nameref{sec:methods}).
        \textbf{d},
        Number of atypical eigenstates $n_{\mathrm{atyp}}$ in the
        largest subspace versus $\phi$ ($N = 15$),
        normalized by $2^N$
        (logarithmic vertical axis; exact zeros omitted).
        \textbf{e}--\textbf{h},
        Extrapolations of \textbf{a}--\textbf{d} in $1/N$
        for $N = 10$--$15$.
        \textbf{e},
        Natural flippability, the mean $\phi$ of the ensemble of
        \textbf{a}.
        \textbf{f},
        Largest-subspace fraction at $\phi = 1$ from panel \textbf{b}.
        \textbf{g},
        GOE-confident flippability $\phi_{\langle r \rangle}$ from
        panel \textbf{c}.
        \textbf{h},
        Largest-subspace $n_{\mathrm{atyp}}/2^N$ at $\phi = 1$ from
        panel \textbf{d}.
        Error bars are standard deviations over $100$ networks per
        flippability level (\textbf{b}--\textbf{d}),
        $10\,000$ draws (\textbf{e}),
        $300$ networks at $\phi = 1$ (\textbf{f}, \textbf{h})
        and $2000$ bootstrap resamples of the networks (\textbf{g}).
        Dashed lines are weighted linear fits;
        crosses (\textbf{e}, \textbf{g}) mark the $N \to \infty$
        intercepts with the fit's standard error,
        and arrows (\textbf{f}, \textbf{h}) mark where the fit reaches
        its bound.
    }
    \label{fig:fig-02}
\end{figure*}

\section*{Static signatures of flippability}
We investigate the static features of the Hamiltonian of
equation~\eqref{eq:Hamiltonian}.
The probability of finding the flippability $\phi$ in the random
ensemble of Hamiltonians is shown in Fig.~\ref{fig:fig-02}a.
The probability distribution has a single peak at $\phi \approx 0.5$,
and low or high flippability is rare.
Randomly generated realizations of the Hamiltonian are thus biased toward
mild flippability.
We therefore generate the Hamiltonians by a balanced protocol of the
flippability $\phi$ to cover the whole range of $\phi \in [0, 1]$ uniformly,
by drawing networks at each flippability level
and assigning the longitudinal fields so that the desired
flippability is achieved
(\nameref{sec:flippability-balanced-protocol} in \nameref{sec:methods}).
In this way, we can mitigate the sampling bias of the random ensemble.

In Fig.~\ref{fig:fig-02}b, we show the fraction of the largest
kinetically accessible subspace in the largest equienergy sector.
As the flippability increases, the fraction increases, meaning that the
equienergy sector becomes less fragmented:
as flippability increases, the largest kinetically accessible subspace tends to
encompass a greater fraction of Fock states within its equienergy sector.

Figure~\ref{fig:fig-02}c shows the mean level-spacing ratio
$\langle r \rangle$ of the eigenenergy spectrum,
which diagnoses the nonintegrability/chaoticity of the
system~\cite{Atas2013}
(\nameref{sec:energy-level-statistics} in \nameref{sec:methods}).
In the Poisson (integrable) limit, $\langle r \rangle \approx 0.386$,
and in the Gaussian orthogonal ensemble (GOE) limit,
$\langle r \rangle \approx 0.531$.
We analyze both the equienergy sector and the largest
kinetically accessible subspace in it.
The mean level-spacing ratio of the whole sector stays at the
Poisson limit up to flippability $\phi \approx 0.8$
and rises to the GOE limit around $\phi \approx 0.9$,
where the whole sector appears to be nonintegrable.
Note that this behaviour coincides with the growth of the largest subspace
fraction, cf. Fig.~\ref{fig:fig-02}b.
If we restrict the analysis to the largest kinetically accessible subspace,
the mean level-spacing ratio rises from flippability $\phi \approx 0.4$,
comes close to the GOE limit at
$\phi_{\langle r \rangle} \approx 0.67$,
and converges to the GOE limit as the flippability increases
(Fig.~\ref{fig:fig-02}c).

We then investigate the number of eigenstates with atypical
participation entropy
(\nameref{sec:participation-entropy} in \nameref{sec:methods}).
Participation entropy quantifies how many Fock states are involved in a
given ket vector,
diagnosing the localization of the state in the Fock-space
landscape~\cite{Roy2025}.
We define an eigenstate as atypical if its participation
entropy lies below the mean of the largest subspace by more than $3$
standard deviations.
In Fig.~\ref{fig:fig-02}d, we show the number of atypical
eigenstates, $n_{\mathrm{atyp}}$,
normalized by the dimension of the Hilbert space $2^N$.
As the flippability increases, the number of atypical eigenstates
increases exponentially,
by more than two orders of magnitude from $\phi \approx 0.4$ to
$\phi = 1$.
This behaviour indicates that although most states are typical in
terms of the participation entropy,
there are a few atypical eigenstates.
As shown in Fig.~\ref{fig:fig-02}c, the system becomes nonintegrable
as the flippability increases,
but there are a few atypical eigenstates in this regime,
reminiscent of the quantum many-body scar states~\cite{Turner2018,Turner2018a}.

We validate our findings in the thermodynamic limit by extrapolating
the results of Fig.~\ref{fig:fig-02}a--d against $1/N$
for $N \in \{10, 11, 12, 13, 14, 15\}$.
The natural flippability, which is the mean flippability from
Fig.~\ref{fig:fig-02}a at each $N$,
is shown in Fig.~\ref{fig:fig-02}e.
We find that the natural flippability converges to $\phi \approx 0.5$ in
the thermodynamic limit, indicating that the distribution of the
realizations along the flippability is not influenced by the system size.
The largest-subspace fraction at $\phi = 1$ from Fig.~\ref{fig:fig-02}b
for various $N$ is shown in Fig.~\ref{fig:fig-02}f.
The fraction is expected to saturate at $1$ around $N \approx 24$;
thus, in the thermodynamic limit,
the largest subspace is expected to dominate the equienergy sector.
The GOE-confident flippability $\phi_{\langle r \rangle}$ from
Fig.~\ref{fig:fig-02}c for various $N$ is shown in Fig.~\ref{fig:fig-02}g.
As we increase the system size, the GOE-confident flippability gets smaller,
and its extrapolation falls below zero
($\phi_{\langle r \rangle} \to -0.14 \pm 0.09$),
so that in the thermodynamic limit the largest subspace is
expected to be nonintegrable at any nonzero flippability.
The largest-subspace $n_{\mathrm{atyp}}/2^N$ at $\phi = 1$ from
Fig.~\ref{fig:fig-02}d for various $N$ is shown in Fig.~\ref{fig:fig-02}h.
The number is expected to saturate at $0$ around $N \approx 247$;
thus, in the thermodynamic limit,
the largest subspace is expected to have an exponentially small number of
atypical eigenstates,
which is consistent with the expectation from the GOE limit.

\begin{figure*}[tb]
    \centering
    \includegraphics{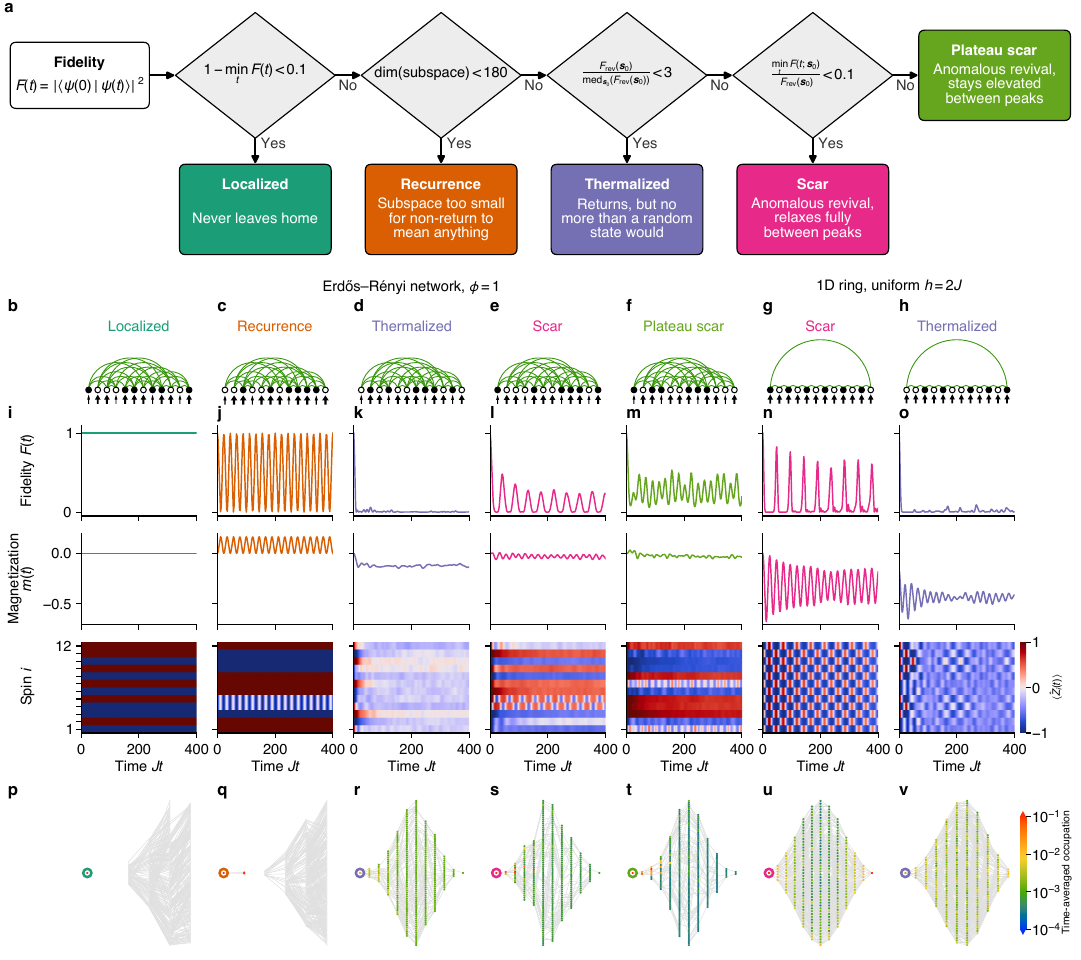}
    \bcaption{
        Examples of dynamical features of the Hamiltonian.
    }{
        \textbf{a},
        Decision flowchart classifying a Fock state of the largest
        equienergy sector into one of five dynamical classes from its own
        fidelity $F(t)$
        (see \nameref{sec:classification} in \nameref{sec:methods}).
        \textbf{b}--\textbf{h},
        Initial Fock state and a realization of network and external field.
        The drawing rule is the same as in Fig.~\ref{fig:fig-01}a.
        \textbf{i}--\textbf{o},
        Time evolution of the fidelity
        $F(t) = |\langle \psi(0) | \psi(t) \rangle|^2$ (top),
        the magnetization per spin
        $m(t) = (1/N) \sum_i \langle \hat{Z}_i(t) \rangle$ (middle),
        and the spin expectation value
        $\langle \hat{Z}_i(t) \rangle$ (bottom).
        \textbf{p}--\textbf{v},
        The Fock-space landscapes coloured by the time-averaged
        occupation probability.
        Each landscape shows only the $322$ most-occupied Fock states,
        the dimension of the ring sector in \textbf{u}--\textbf{v}.
        The initial state (circle) is on the left,
        and the Fock states are ordered from left to right by their
        Hamming distance from it.
        Grey lines connect kinetically accessible Fock states
        (one spin flip apart with equal classical energy).
        Panels \textbf{b}--\textbf{v} are created from one realization
        of an interaction network and longitudinal fields
        ($N = 12$) at $\phi = 1$,
        with five initial Fock states of zero total magnetization,
        $\sum_i s_i = 0$
        (columns \textbf{b}--\textbf{f}, \textbf{i}--\textbf{m},
            \textbf{p}--\textbf{t}:
        localized, recurrence, thermalized, scar, plateau scar)
        and, to their right, the one-dimensional ferromagnetic ring with
        uniform longitudinal field $h_i = 2J$ for all $i$
        (columns \textbf{g}--\textbf{h}, \textbf{n}--\textbf{o},
            \textbf{u}--\textbf{v}:
            staggered scar and a thermalizing Fock state with five up spins
            in the same sector;
        no Fock state with six up spins thermalizes in this sector).
        The network selection rule is given in
        \nameref{sec:example-networks}
        in \nameref{sec:methods}.
    }
    \label{fig:fig-03}
\end{figure*}

\section*{Dynamics at high flippability}
In the previous section, we showed that the flippability manifests itself as
a powerful indicator to predict qualitative differences among Hamiltonians
with random interaction networks and random longitudinal fields.
In this section, we further validate our findings by exploring the
dynamical features of the Hamiltonian.

Our interest in the quantum many-body scar states motivates us to
investigate the initial-state memory of the system.
Initializing the system in a Fock state,
$
\ket{\psi(0)} = \ket{\bm{s}_0}
$,
we run the time evolution of the system following the Schr\"odinger equation,
with $\hat{\mathcal{H}}$ given by the effective Hamiltonian of
equation~\eqref{eq:approximated-Hamiltonian} restricted to the largest
equienergy sector,
as in all numerical results of this work
\begin{equation}
    \mathrm{i}
    \frac{\partial}{\partial t}
    \ket{\psi(t)}
    =
    \hat{\mathcal{H}}
    \ket{\psi(t)}
    .
\end{equation}
From the time evolution of the ket vector,
we compute the time evolution of the fidelity
\begin{equation}
    F(t; \bm{s}_0)
    \coloneqq
    |\! \braket{\psi(0) | \psi(t)} \!|^2
    =
    |\! \braket{\bm{s}_0 | \psi(t)} \!|^2
    \label{eq:fidelity}
    .
\end{equation}
The fidelity quantifies how similar the time-evolved state $\ket{\psi(t)}$ is
to the initial state $\ket{\psi(0)} = \ket{\bm{s}_0}$,
diagnosing the initial-state memory of the system.
Using the fidelity,
the initial Fock states are classified into five dynamical classes:
localized, recurrence, thermalized, scar, and plateau scar,
following the flowchart of Fig.~\ref{fig:fig-03}a
(\nameref{sec:classification} in \nameref{sec:methods}).
The localized states are the Fock states that do not evolve in time,
which can be classified by thresholding the minimum of fidelity.
The recurrence states are the Fock states that recur to the initial one
due to the small size of the kinetically accessible subspace.
The recurrence states can be classified by examining the dimension of
the kinetically accessible subspace.
The thermalized states are the Fock states that lose the
initial-state memory;
their fidelity decays and fluctuates around zero.
By checking the ratio between the maximum of the revival and the median of
that across other Fock states,
the thermalized states can be classified.
The scar states exhibit periodic initial-state revival in the fidelity,
which can be classified by thresholding the ratio between the minimum
and maximum of the fidelity.
In other words, if the minimum of the fidelity is smaller than
10\% of the maximum of the revival,
the Fock state is classified as a scar state.
If not, the minimum of the fidelity has a magnitude comparable to the
maximum of the revival.
We call such states plateau-scar states to emphasize their difference
from the scar states originally proposed in the
seminal works~\cite{Turner2018,Turner2018a}.

To demonstrate the classification of the Fock states into five dynamical
classes, we show examples of the time evolution of the fidelity and
the spin expectation value in Fig.~\ref{fig:fig-03}i--o.
We choose a realization of the Hamiltonian with $N = 12$ and $\phi = 1$
(\nameref{sec:example-networks} in \nameref{sec:methods}),
and simulate the time evolution starting from the five representative
Fock states of each dynamical class.
As a reference, we also show the time evolution of the fidelity and
the spin expectation value of the one-dimensional ferromagnetic ring
with uniform longitudinal field $h_i = 2J$,
which corresponds to the PXP model~\cite{Yoshinaga2022},
exhibiting scar states if the initial state is the staggered/N\'eel Fock state.
In Fig.~\ref{fig:fig-03}b--h,
we show the initial Fock states and the corresponding interaction
network and the longitudinal fields.
Below them, in Fig.~\ref{fig:fig-03}i--o,
we show the time evolution of the fidelity and the spin expectation value.
The localized state in Fig.~\ref{fig:fig-03}i shows that the fidelity
remains constant,
and the spin expectation value does not change in time.
The recurrence state in Fig.~\ref{fig:fig-03}j shows that the
fidelity exhibits periodic revival and the spin expectation values
of two spins oscillate in time,
indicating that the initial state in Fig.~\ref{fig:fig-03}c is
kinetically connected with few other Fock states.
The thermalized state in Fig.~\ref{fig:fig-03}k shows that the fidelity
decays to zero and fluctuates around it,
and the spin expectation values lose their initial polarized values.
The scar state in Fig.~\ref{fig:fig-03}l shows that the fidelity
exhibits periodic and decaying revival,
and the spin expectation values of three spins strongly oscillate and
revive to their initial values in time,
while the other spins oscillate more weakly.
The plateau-scar state in Fig.~\ref{fig:fig-03}m shows that the fidelity
decays to a plateau value and fluctuates around it,
and the spin expectation values decay to a plateau value and
fluctuate around it as well.
Notably, all initial Fock states of Fig.~\ref{fig:fig-03}i--m
have the same classical energy and zero magnetization per spin,
$m(0) = (1/N) \sum_i \langle \hat{Z}_i(0) \rangle = 0$,
yet their time evolutions differ,
indicating that the microscopic details of the initial Fock state,
the arrangement of its up spins on the network,
alter the relaxation dynamics of the macroscopic
observable~\cite{Yoshinaga2022}.
Those rich behaviours of the random Hamiltonian of
equation~\eqref{eq:Hamiltonian} are compared with those of the
ring in Fig.~\ref{fig:fig-03}n--o.
In Fig.~\ref{fig:fig-03}n, the staggered scar state shows that the fidelity
exhibits periodic revival as in the random system,
but the time evolution of the spin expectation value is highly coherent and
involves strong oscillations of all spins~\cite{Choi2019}.
The thermalizing Fock state in Fig.~\ref{fig:fig-03}o shows the same
behaviour as the random system in Fig.~\ref{fig:fig-03}k.

Those dynamical behaviours are further examined in the Fock-space landscape
visualized in Fig.~\ref{fig:fig-03}p--v.
We visualize the Fock-space landscape by drawing the Fock states
in the order
of their Hamming distance from the initial Fock state,
and colour the Fock states by the time-averaged occupation probability.
As predicted from Fig.~\ref{fig:fig-03}i,
the localized initial Fock state is kinetically isolated from other
Fock states in the landscape (Fig.~\ref{fig:fig-03}p).
Similar to the localized initial Fock state,
the recurrence initial Fock state has few connected Fock states in
the landscape (Fig.~\ref{fig:fig-03}q), and the probability
distribution on the landscape is localized.
On the other hand, the thermalized initial Fock state has many connected
Fock states in the landscape (Fig.~\ref{fig:fig-03}r),
and the probability distribution on the landscape is widely spread
over the landscape.
The scar state also has many connected Fock states in the landscape
(Fig.~\ref{fig:fig-03}s),
but the probability distribution on the landscape is
localized on the initial state and states close to it,
and the probability is less spread over the landscape than the
thermalized state.
The plateau-scar state also has many connected Fock states in the landscape,
but again the probability distribution on the landscape is localized
on the initial state and states close to it (Fig.~\ref{fig:fig-03}t).
Compared to the scar state, the probability distribution of the
plateau-scar state is more spread over the landscape.
Those behaviours are compared with those of the ring in
Fig.~\ref{fig:fig-03}u--v.
The scar state in the ring has fewer connected Fock states
than that of the random system (Fig.~\ref{fig:fig-03}u),
and the structure of the landscape is known to have more regularity
(Fibonacci chain)~\cite{Turner2018,Turner2018a} than that of the
random system.
The probability distribution of the scar state in the ring
is also localized,
but unlike the random system, it is bipolar:
the probability is localized around the initial state and its all-spin-flipped
state.
The ring symmetry and the Fibonacci-chain structure of the landscape
are the source of the system-wide coherent oscillation in the fidelity
and the spin expectation value.
The behaviour of the thermalizing Fock state on the landscape is
qualitatively similar to that of the random system (Fig.~\ref{fig:fig-03}v).

In Fig.~\ref{fig:fig-03}p--v,
we only consider the time-averaged occupation probability on the
Fock-space landscape,
but our discussion remains valid if we consider the instantaneous time
evolution of the occupation probability on the landscape
(Supplementary Movie 1).
The qualitative difference between the scar state of the random
Hamiltonian and that of the ring is also observed
(Supplementary Movie 2).
If we marginalize the probability distribution onto the three spins
with the largest oscillation amplitude,
the scar state flips spins 4, 12, and 5 in sequence,
as the ring does
(Supplementary Movie 3).
The Hamiltonian of equation~\eqref{eq:Hamiltonian} thus has a scar
state with sequential, in addition to collective, spin flips.

Together, those results demonstrate that the random Hamiltonian of
equation~\eqref{eq:Hamiltonian} exhibits richer dynamical behaviours
than the ring,
and their scar states have qualitatively different behaviours from each other.
In addition, the Hamiltonian of equation~\eqref{eq:Hamiltonian} has
plateau-scar states, which are not observed in the ring.

\begin{figure*}[tb]
    \centering
    \includegraphics{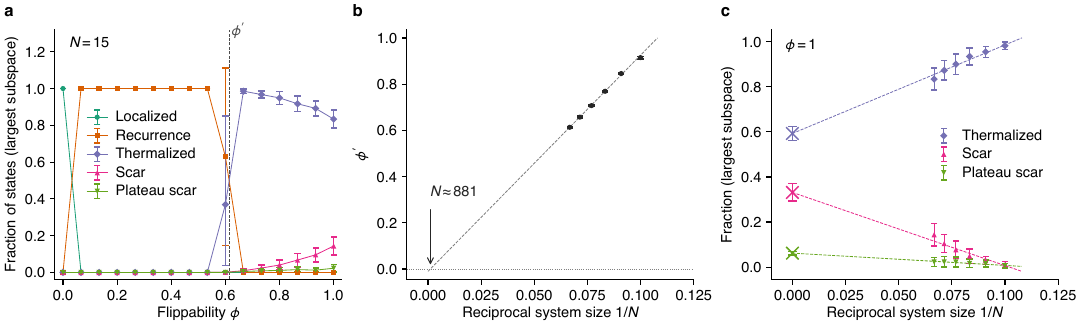}
    \bcaption{
        Fractions of classified states and their
        thermodynamic-limit extrapolations.
    }{
        \textbf{a},
        Fractions of the five dynamical classes in the largest subspace
        versus flippability $\phi$ ($N = 15$).
        The dashed vertical line indicates the thermalization-dominant
        flippability $\phi^\prime$ where the recurrence and
        thermalized fractions cross.
        \textbf{b},
        Thermalization-dominant flippability $\phi^\prime$
        versus $1/N$ for
        $N \in \{10, 11, 12, 13, 14, 15\}$,
        bootstrapped over networks.
        The thermalization-dominant flippability $\phi^\prime$ is
        expected to reach zero at $N \approx 881$.
        \textbf{c},
        Fractions of the thermalized, scar and plateau-scar
        classes versus $1/N$ at $\phi = 1$.
        The same symbols and colours as in panel \textbf{a} are used.
        Crosses indicate the $N \to \infty$ intercepts.
        Error bars show the standard deviation:
        in panel \textbf{a}, over $100$
        networks per flippability level;
        in \textbf{b}, over $2000$ bootstrap resamples of the networks;
        and in \textbf{c}, over the networks at $\phi = 1$
        ($200$ for $N \leq 13$ and $100$ for $N \in \{14, 15\}$).
        Dashed lines are weighted linear fits in $1/N$;
        crosses carry the standard error of the fit,
        and arrows mark where a fit reaches its bound.
        The sampling and extrapolation procedure is given in
        \nameref{sec:class-fraction-statistics} in \nameref{sec:methods}.
    }
    \label{fig:fig-04}
\end{figure*}

We then examine how the class fractions change with the flippability.
In Fig.~\ref{fig:fig-04}a, we show the fractions of the five dynamical
classes in the largest kinetically accessible subspace as a function of
flippability $\phi$.
At $\phi=0$, the initial Fock states are dominated by localized states.
This is expected from Figs.~\ref{fig:fig-03}b, i and p:
the initial Fock state is kinetically isolated from other Fock states.
For $\phi > 0$, the class fraction is dominated by the
recurrence state until $\phi = \phi^\prime \approx 0.6$,
and then the dominance switches to the thermalized state.
This reminds us of the results of Fig.~\ref{fig:fig-02}c:
the largest subspace becomes GOE-like around
$\phi_{\langle r \rangle} \approx 0.67$,
close to $\phi^\prime$,
indicating that the Fock-state-initialized dynamics is related to the
level statistics of the largest subspace.
Although the thermalized state dominates
in the regime of $\phi \gtrsim \phi^\prime$,
its fraction decreases as the flippability increases;
the scar and plateau-scar states acquire the remaining fraction as the
flippability increases.

We then extrapolate these class-fraction results to the thermodynamic limit.
We show, in Fig.~\ref{fig:fig-04}b, the thermalization-dominant flippability
$\phi^\prime$,
where the dominance of the thermalized state begins.
As the system size increases, the thermalization-dominant
flippability decreases,
and around $N \approx 881$, it is expected to converge to zero,
meaning that thermalized states dominate the initial Fock states
in the thermodynamic limit regardless of the flippability.
However, a fraction of scar and plateau-scar states remains in
the thermodynamic limit.
In Fig.~\ref{fig:fig-04}c,
we show how the fraction of the thermalized, scar and plateau-scar
classes depends on the system size at $\phi = 1$.
As the system size increases, the fraction of the thermalized state
decreases, while the fraction of the scar and plateau-scar states increases.
Thus, maximally flippable Hamiltonians are expected to have a finite
fraction of scar and plateau-scar states in the thermodynamic limit.

\section*{Conclusions}
Recent experimental advances have enabled the realization of arbitrary
interaction networks in quantum simulators.
Understanding how the network structure shapes quantum many-body
dynamics remains an open challenge.
In this study, we investigated random Erd\H{o}s--R\'enyi Ising networks
in the weak transverse field,
and identified flippability---the fraction of spins that can flip
without changing the corresponding classical energy---as a predictor
of quantum many-body scar states from the parameters of the Hamiltonian:
the interaction network and longitudinal fields.
Our phenomenological laws of flippability are reminiscent of Euler's
approach to the problem of the Seven Bridges of Königsberg:
we can predict the nontrivial dynamical behaviour by simply counting
the number of interactions.

We have shown that the flippability predicts the fragmentation of the
equienergy sector,
the nonintegrability of the largest kinetically accessible subspace,
and the number of atypical eigenstates in the largest subspace.
Systems with high flippability exhibit diverse dynamical
behaviours, including the localized, recurrence, thermalized, scar,
and plateau-scar states.
Although the fidelity revival behaviour is similar to that of the PXP model,
the scar state of the random system is less coherent than that of the
PXP model~\cite{Choi2019}.
In the thermodynamic limit, systems with any flippability become
nonintegrable, consistent with an earlier study~\cite{Grabarits2025}.
Nevertheless, systems with the maximum flippability are expected to
have a finite fraction of scar and plateau-scar states in the
thermodynamic limit,
showcasing the predictability of the flippability for the
nonequilibrium dynamics of the system.
We confirm that for networks with antiferromagnetic
interactions, $\bm{J} \in \{-J, 0, +J\}^{N\times N}$,
i.e., the $\pm J$ ensemble,
these findings are qualitatively the same
(Extended Data Figs.~\ref{fig:fig-s02}--\ref{fig:fig-s04}).

Our findings provide experimentally verifiable predictions:
tuning the local longitudinal field $h_i$ changes the floppy condition
and hence the flippability,
allowing direct control over scar formation in programmable
quantum simulators.
This work contributes to the emerging field at the intersection
of quantum physics and network science~\cite{Nokkala2024} and
illustrates how the structure of interactions generates new
many-body phenomena.
As graphene and diamond symbolically embody,
even if the elements of the system are the same,
the structure of the interaction network leads to totally different
properties of the system.
Our work highlights the importance of the interaction network
structure---even if it seems random---in the nonequilibrium
dynamics of quantum many-body systems.

\bibliography{references}

\clearpage

\section*{Methods}
\label{sec:methods}

\subsection*{Random network generation}
\label{sec:random-network-generation}
We generate random Erd\H{o}s--R\'enyi networks on $N$ vertices,
in which each of the $N(N-1)/2$ vertex pairs is joined by an edge
independently~\cite{Gilbert1959} with probability $p_\mathrm{ER} = 0.5$,
and every edge carries the weight $+J$.
Erd\H{o}s and R\'enyi showed that the network becomes connected at
$p_\mathrm{ER} \simeq \ln(N)/N$~\cite{Erdos1959}:
for $p_\mathrm{ER} = [\ln(N) + c]/N$,
the probability that the network is connected approaches
$\exp[-\exp(-c)]$.
They derived this threshold for networks with a fixed number of edges,
and the same threshold holds for the model above.
Below this threshold,
the network typically splits into independent subsystems without
interaction.
In the opposite limit $p_\mathrm{ER} \to 1$,
every spin interacts with every other,
the system becomes mean-field-like,
and the network structure plays a lesser role
(see Supplementary Information for a discussion of the all-to-all model).
We choose $p_\mathrm{ER} = 0.5$ between these two limits.
It lies above the connectivity threshold for every system size we
consider [$\ln(N)/N \leq 0.23$ for $N \geq 10$].
We discard disconnected realizations and redraw.
For the $\pm J$ ensemble
(results in Extended Data Figs.~\ref{fig:fig-s02}--\ref{fig:fig-s04}),
each nonzero edge is assigned $J_{i,j} \in \{+J, -J\}$
independently with equal probability.

\subsection*{The flippability-balanced protocol}
\label{sec:flippability-balanced-protocol}
For Figs.~\ref{fig:fig-02} and~\ref{fig:fig-04},
longitudinal fields are assigned using the balanced
flippability protocol:
for each of the $N + 1$ flippability levels
$\phi = a / N$, $a \in \{0, 1, 2, 3, \ldots, N\}$,
we draw independent networks
and, for each network, randomly select $a$ spins to be flippable
and assign their $h_i$ values accordingly.
This ensures uniform coverage of flippability independently
of the network structure.

\subsection*{Energy level statistics}
\label{sec:energy-level-statistics}
To quantify spectral chaos, we compute the mean consecutive
level-spacing ratio~\cite{Atas2013}
\begin{align}
    \langle r \rangle
    &\coloneqq
    \frac{1}{M-2}
    \sum_{\mu=1}^{M-2}
    \frac{
        \min(\varDelta_\mu, \varDelta_{\mu+1})
    }{
        \max(\varDelta_\mu, \varDelta_{\mu+1})
    }
    ,
\end{align}
where $\varDelta_\mu \coloneqq E_{\mu+1} - E_\mu$ is the $\mu$th
level spacing,
and eigenvalues $\{E_\mu\}$ are sorted in ascending order.
Here, the average is taken over the sector/subspace of interest.
The factor $M$ is the number of levels retained in the chosen
sector or subspace, as described below.
This ratio converges to $\langle r \rangle \approx 0.386$ for integrable
(Poisson) spectra and $\langle r \rangle \approx 0.531$ for
nonintegrable/chaotic (GOE) spectra~\cite{Atas2013}.

We define the GOE-confident flippability $\phi_{\langle r \rangle}$
as the flippability at which the largest subspace first becomes
chaotic within a fixed tolerance.
At each flippability level, we average $\langle r \rangle$ of the largest
subspace over the networks whose largest subspace has at least $20$ levels,
and keep the level only if at least half of its networks,
and at least $5$,
qualify;
where only a minority qualifies,
their largest subspaces barely exceed $20$ levels,
and $\langle r \rangle$ falls below the Poisson value as a finite-size
artefact.
Let $\overline{\langle r \rangle}$ be this average.
Then $\phi_{\langle r \rangle}$ is the lowest level that satisfies
\begin{align}
    \frac{
        | 0.531 - \overline{\langle r \rangle} |
    }{
        0.531 - 0.386
    }
    < 0.1
    ,
\end{align}
that is, $\overline{\langle r \rangle}$ lies within $10\%$ of the
Poisson--GOE range from the GOE value
($\overline{\langle r \rangle} \gtrsim 0.517$).
We do not interpolate between the discrete flippability levels.
To estimate the uncertainty,
we resample the networks at each level with replacement,
repeat the procedure for $2000$ resamples,
and report the mean and the standard deviation of the resulting
$\phi_{\langle r \rangle}$.

The spectrum of an equienergy sector has a trivial degeneracy.
Each hop of equation~\eqref{eq:approximated-Hamiltonian} flips a single
spin,
so the hopping graph is bipartite,
while the diagonal elements equal the constant sector energy $E_0$;
the spectrum is therefore symmetric about $E_0$,
with degenerate zero modes at $E_0$,
and identical fragments of a disconnected sector share their
eigenvalues.
To lift these degeneracies,
we add a small longitudinal field $\sum_i \epsilon_i \hat{Z}_i$
to the sector Hamiltonian,
where $\epsilon_i$ is drawn once per realization from the uniform
distribution on $[-0.1\varGamma, 0.1\varGamma]$.
The field is diagonal in the Fock basis,
so it does not couple different kinetically accessible subspaces.

\subsection*{Participation entropy}
\label{sec:participation-entropy}
To quantify the localization of the ket in the Fock-space landscape,
we compute the participation entropy~\cite{Roy2025}.
The participation entropy is defined as
\begin{equation}
    S_\alpha(\ket{\psi})
    \coloneqq
    -
    \frac{1}{\alpha-1}
    \ln
    R_\alpha(\ket{\psi})
    ,
\end{equation}
where $\alpha$ is the order of the participation entropy and
\begin{equation}
    R_\alpha(\ket{\psi})
    \coloneqq
    \sum_{\bm{s}}
    p^\alpha (\bm{s})
    =
    \sum_{\bm{s}}
    |c(\bm{s})|^{2 \alpha}
    ,
\end{equation}
is the $\alpha$th inverse participation ratio, and
$c(\bm{s}) \coloneqq \braket{\bm{s} | \psi}$ is the coefficient of
the Fock state $\ket{\bm{s}}$ in the wave function $\ket{\psi}$.

For $\alpha = 2$, the participation entropy is known as the Rényi-2
participation entropy.
For a GOE spectrum, its average over the eigenstates of a subspace of
dimension $M$ is
\begin{equation}
    \overline{S_2 (\ket{\psi})}
    =
    \ln (M)
    -
    \ln (3)
    +
    \frac{10}{3M}
    +
    O(M^{-2})
    \label{eq:s2-eth}
    ,
\end{equation}
whose leading terms are known~\cite{Backer2019};
we derive the finite-size correction $10/(3M)$ in the Supplementary
Information.
Here, $M$ is the dimension of the sector or subspace.
Notably, the participation entropy is independent of the
eigenstate's energy
unlike the entanglement entropy,
and we can systematically detect the atypical eigenstates in the
sector by comparing the participation entropy.

We define an eigenstate as atypical~\cite{Beugeling2015}
when its participation entropy falls more than $3$ standard
deviations below the average of that in the sector,
\begin{align}
    &\ket{\psi} \text{ is atypical if }
    \nonumber
    \\&S_2 (\ket{\psi})
    -
    \overline{
        S_2 (\ket{\psi})
    }
    <
    -3
    \,
    \operatorname{std}\bm{(}S_2(\ket{\psi})\bm{)}
    \label{eq:atypical-criterion}
    ,
\end{align}
where the average and standard deviation are calculated over the
sector of interest.
We denote by $n_{\mathrm{atyp}}$ the number of eigenstates in a sector
satisfying this criterion.
Within a degenerate eigenspace,
such as the zero modes at $E_0$,
the eigenstates are not unique,
and neither is their participation entropy.
We therefore compute the eigenstates with the small longitudinal field
$\epsilon_i \in [-0.1\varGamma, 0.1\varGamma]$ of
\nameref{sec:energy-level-statistics},
which lifts the degeneracies,
and diagonalize each kinetically accessible subspace separately.
In Fig.~\ref{fig:fig-02}d,
the average and standard deviation are taken over the eigenstates of
the largest kinetically accessible subspace,
and $n_{\mathrm{atyp}}$ counts the atypical eigenstates in it.
We do not use equation~\eqref{eq:s2-eth} as the reference,
because the eigenstates of the kinetically accessible subspaces deviate
from it systematically
(Extended Data Fig.~\ref{fig:fig-s05}a--c).
Instead, we define the average of the participation entropy as the
typical value of the participation entropy in the sector of interest.
See the Supplementary Information and Extended Data
Figs.~\ref{fig:fig-s05}--\ref{fig:fig-s07} for the comparison between the
participation entropy and the multipartite entanglement.

\subsection*{Five-class classification of Fock states}
\label{sec:classification}
An initial Fock state
$
\ket{\psi(0)}
=
\ket{\bm{s}_0}
$
is classified into one of five dynamical classes from
its own fidelity time evolution
$
F(t; \bm{s}_0)
=
|\! \braket{\psi(0) | \psi(t)} \!|^2
=
|\! \braket{\bm{s}_0 | \psi(t)} \!|^2
$,
by applying the tests in the order of Fig.~\ref{fig:fig-03}a.
We define the revival value as
\begin{align}
    F_\mathrm{rev}(\bm{s}_0)
    \coloneqq
    \max_{t_\mathrm{relax} \leq t \leq \tau}
    F(t; \bm{s}_0),
\end{align}
the largest fidelity reached after the relaxation time
$t_\mathrm{relax} = 50/J$,
by which the thermalized states have relaxed
(Extended Data Fig.~\ref{fig:fig-s07}),
within a time window of length $\tau = 2000/J$.
Starting the search at $t_\mathrm{relax}$ excludes the initial decay
of the fidelity
and its early ringing,
which would otherwise count as a return.
A state that never leaves home is caught by the localization test
before its revival is examined.
The four numerical thresholds that appear in the flowchart
are fixed once
and for all; each is justified below and calibrated in
the Supplementary Information (Extended Data Fig.~\ref{fig:fig-s08}).

\paragraph*{Localization threshold.}
A state is \emph{localized} if it never leaves,
$1 - \min_t F(t; \bm{s}_0) < 0.1$, or equivalently, if the fidelity
never falls below
$0.9$.
States that fail this test are asked whether they returned.

\paragraph*{Recurrence threshold.}
We then ask whether a state trivially returns to the initial state or not.
A fidelity revival has no absolute scale,
because how far a state returns depends on how large a space it explores.
To calibrate the dimension dependence,
we evolved random basis states under random GOE Hamiltonian matrices.
Each GOE spectrum is rescaled to the median bandwidth of the real
subspaces of the same dimension $D$
(Supplementary Information).
The median revival of this null exceeds $0.1$,
the same scale as the localization threshold,
for $D \lesssim 180$.
In such a subspace a revival carries no information beyond its size,
and we call the state a recurrence.

\paragraph*{Thermalization threshold.}
Above the recurrence regime, we compare a state to the other states of
its own subspace, dividing its revival by their median and calling the
ratio the anomaly,
\begin{align}
    \frac{
        F_\mathrm{rev}(\bm{s}_0)
    }
    {
        \med_{\bm{s}_0}\bm{(}F_\mathrm{rev}(\bm{s}_0)\bm{)}
    }\,,
    \label{eq:thermalization-threshold}
\end{align}
where
$\med_{\bm{s}_0}\bm{(}F_\mathrm{rev}(\bm{s}_0)\bm{)}$
is the median of
$F_\mathrm{rev}(\bm{s}_0)$ over all initial Fock states $\{
\ket{\bm{s}_0} \}$
sharing the same subspace dimension, $D \coloneqq \dim (\mathrm{subspace})$,
within the same realization.
We use this median rather than the GOE null as the reference,
because the subspaces of equation~\eqref{eq:Hamiltonian} are sparse
graphs that recur more than random matrices;
against the null,
the typical state of a large subspace would itself count as anomalous
(Extended Data Fig.~\ref{fig:fig-s08}c).
We call a state anomalous if its anomaly is at least $3$,
well above the $99$th percentile of the null,
which lies at most $1.8$ times above its median,
and thermalized otherwise.

\paragraph*{Scar threshold.}
An anomalous state is further split by the floor ratio
\begin{equation}
    \frac{
        \min_t F(t; \bm{s}_0)
    }
    {
        F_\mathrm{rev} (\bm{s}_0)
    }
    .
\end{equation}
A value close to $0$ means the state fully relaxes and then revives
(an oscillating scar);
a value close to $1$ means it settles at a persistently elevated
fidelity after the first collapse and never fully relaxes again
(a \emph{plateau} scar).
Across the scar population this ratio is smooth and unimodal with no
natural gap, so the threshold
$\min_t F(t; \bm{s}_0) / F_\mathrm{rev} (\bm{s}_0) = 0.1$
is a convention on a continuum, chosen to match the same $0.1$ scale as
the localized and recurrence thresholds:
the floor after collapse must stay at least a tenth of the revival
peak to count as a plateau.

\subsection*{Example networks in Fig.~\ref{fig:fig-03}}
\label{sec:example-networks}
The ferromagnetic block of Fig.~\ref{fig:fig-03}b--v shows one
realization at $N = 12$ and $\phi = 1$.
All initial Fock states have zero magnetization, $\sum_i s_i = 0$,
and lie in the largest equienergy sector,
so the five columns differ only in where the up spins sit on the
network.
Among the first $720$ realizations,
a network qualifies if this sector contains states of all five classes,
including a recurrence in which exactly two spins move
and a scar finalist:
an oscillating scar with excursion
$\max_t \langle \hat{Z}_i(t) \rangle - \min_t \langle \hat{Z}_i(t) \rangle
> 0.6$ on at least nine spins,
at least five resolved peaks after the collapse,
$F_\mathrm{rev} \geq 0.4$,
and at least a fifth of the window below $F(t) = 0.05$.
We show the qualifying network whose scar is most PXP-like:
among the scar states with at least five peaks above $F(t) = 0.2$,
the scar column has none between $F(t) = 0.02$ and $0.2$
and spends the largest fraction of $Jt \leq 400$ below $F(t) = 0.01$.
The thermalized column has the largest overlap with the eigenstate
whose participation entropy is closest to the average over its
subspace;
the plateau-scar column is the plateau scar of highest anomaly;
the recurrence column lies in the smallest subspace with two moving
spins, ties broken by the revival ratio;
and the localized column is a one-dimensional subspace.

The ring block in Fig.~\ref{fig:fig-03} is created from the simulation of
the fixed one-dimensional ferromagnetic ring with
$N = 12$, $h_i = 2J$ for all $i$:
the staggered product state and one thermalizing Fock state with five
up spins in the same sector.
The $\pm J$ analogue (Extended Data Fig.~\ref{fig:fig-s03}) uses the
same qualification
but selects,
among the qualifying networks of the first $60$ realizations that
contain a plateau-scar state of zero magnetization,
the scar candidate of the highest
spectral purity
(the largest fraction of revival AC power in a single Fourier
component),
and shows that candidate in its scar column;
its ring is a Mattis gauge transform of the ferromagnetic ring.

\subsection*{Class-fraction statistics in Fig.~\ref{fig:fig-04}}
\label{sec:class-fraction-statistics}
For each system size $N$ and target flippability $\phi$,
we generated random network realizations
($200$ for $N \leq 13$, $100$ for $N \in \{14, 15\}$)
and classified every Fock state of each realization's largest equienergy
sector by the five-class rule (\nameref{sec:classification}).
The class fractions in Fig.~\ref{fig:fig-04} are computed over the
Fock states of the largest kinetically accessible subspace.
A kinetically accessible subspace of the sector
with more than $800$ Fock states was not evolved in full:
we drew $n_\mathrm{meas} = 100$ of its Fock states at random,
without replacement,
and evolved and classified only those.
Subspaces with at most $800$ Fock states were evolved exhaustively.
Each measured state stands in for $D / n_\mathrm{meas}$ states
of its subspace and is counted with that weight,
where $D$ is the dimension of the subspace
(weight $1$ for an exhaustively evolved one),
so that the reported class fractions remain unbiased;
this sampling binds only close to $\phi = 1$,
where the largest subspace reaches about $6800$ states at $N = 15$.

The thermalization-dominant flippability $\phi^\prime$ at each
$N \in \{10, 11, 12, 13, 14, 15\}$
in Fig.~\ref{fig:fig-04}b is itself a bootstrap estimate:
at each $\phi$ level the networks are resampled with replacement,
a resampled mean recurrence/thermalized fraction curve is built,
and the crossing is located;
repeated $2000$ times,
this gives a mean and standard deviation of $\phi^\prime$ at that $N$.
The resulting $(1/N, \phi^\prime)$ points are fit to a line in $1/N$,
weighted by the bootstrap standard deviation.

\subsection*{Use of large language models}
Claude (Anthropic) was used to assist in refactoring,
writing and documenting the analysis code,
in computing summary values reported in the Methods,
and in editing the manuscript text.
The authors reviewed and verified all code, numerical results and text,
and take full responsibility for the content of this paper.

\subsection*{Data availability}
All data supporting the findings of this study,
including the data shown in every figure,
are available at Zenodo~\cite{Horiike2026}
(https://doi.org/10.5281/zenodo.23429190).

\subsection*{Code availability}
Calculations and visualizations of this work were performed
using open-source
Python~\cite{Rossum2010} libraries:
Matplotlib~\cite{Hunter2007}, NetworkX~\cite{Hagberg2008},
NumPy~\cite{Harris2020},
and SciPy~\cite{Virtanen2020}.
For some figures, the colour maps of
ColorCET~\cite{Kovesi2015} are used.
The code that generates the data, figures, and movies
is available at Zenodo~\cite{Horiike2026}
(https://doi.org/10.5281/zenodo.23429190).

\bigskip
\noindent\footnotesize\textbf{Acknowledgments}
Y.H.~thanks Andrew D. King for the introduction to ref.~\cite{King2016},
and Koki Mizuno for fruitful discussions.

\noindent\textbf{Funding}
Y.H.~was supported by
SPRING Grant No.~JPMJSP2125
``THERS Make New Standards Program for the Next Generation Researchers''
of the Japan Science and Technology Agency,
travel grant No.~M25SN034 by Murata Science and Education Foundation,
the 8th ANRI Fellowship of ANRI Inc.,
the Research Fellowships of the Japan Society for the Promotion of Science
for Young Scientists,
and the KAKENHI Grant No.~JP26KJ1295 of the Japan Society for the
Promotion of Science.
Y.K. was supported by KAKENHI Grant No.~24K00557 of the Japan
Society for the Promotion of Science.

\noindent\textbf{Author contributions}
Y.H.~performed conceptualization, data curation, formal
analysis, validation,
and visualization.
Y.H.~conducted investigation and developed methodology and software.
Y.H.~and Y.K.~acquired funding.
Y.K.~provided computing resources, administered the project,
and supervised the work.
Y.H.~wrote the original draft, and Y.H.~and Y.K.~reviewed and edited
the manuscript.

\noindent\textbf{Competing interests}
The authors declare no competing interests.

\bigskip
\noindent\textbf{Additional information}\\
\noindent\textbf{Correspondence and requests for materials}
should be addressed to Y.H.

\end{document}


\title{
    Supplementary Information for\\
    ``Nonequilibrium quantum states emerging from flippable spins
    in random networks''
}

\author{Yoshiaki Horiike\orcidlink{0009-0000-2010-2598}}
\email{yoshi.h@nagoya-u.jp}
\affiliation{
    Department of Applied Physics, Nagoya University, Nagoya, Japan
}

\author{Yuki Kawaguchi\orcidlink{0000-0003-1668-6484}}
\affiliation{
    Department of Applied Physics, Nagoya University, Nagoya, Japan
}
\affiliation{
    Research Center for Crystalline Materials Engineering,
    Nagoya University,
    Nagoya, Japan
}

\date{17 October 2026}

\begingroup
\let\clearpage\relax
\maketitle
\endgroup
\tableofcontents

\newpage

\section{Validation of the equienergy approximation of the Hamiltonian}
\label{sec:validation-approx-h}
In the main text, the Fock-space Hamiltonian
\begin{equation}
    \hat{\mathcal{H}}
    =
    \sum_{\bm{s}}
    H(\bm{s})
    \ket{\bm{s}} \! \bra{\bm{s}}
    -
    \varGamma
    \sum_{\bm{s}}
    \sum_{k=1}^N
    \ket{\bm{F}_{k} \bm{s}} \! \bra{\bm{s}}
\end{equation}
is approximated, for weak transverse field
$|\varGamma|\ll |J|$,
by retaining only equienergy (floppy) hops,
\begin{equation}
    \hat{\mathcal{H}}
    \approx
    \sum_{\bm{s}}
    H(\bm{s})
    \ket{\bm{s}} \! \bra{\bm{s}}
    -
    \varGamma
    \sum_{\bm{s}}
    \sum_{k=1}^N
    \delta_{H(\bm{s}), H(\bm{F}_{k} \bm{s})}
    \ket{\bm{F}_{k} \bm{s}} \! \bra{\bm{s}}
    ,
    \label{eq:supp-approx-H}
\end{equation}
with the classical energy
\begin{align}
    H(\bm{s})
    &=
    -
    \frac{1}{2}
    \sum_{i=1}^{N}
    \sum_{j=1}^{N}
    s_i
    J_{i,j}
    s_j
    -
    \sum_{i=1}^{N}
    s_i
    h_i
    \\
    &=
    -
    \frac{1}{2}
    \bm{s}^\top \bm{J} \bm{s}
    -
    \bm{s}^\top \bm{h}
\end{align}
as in the main text.
Intuitively, a hop that changes the classical energy is off-resonant:
it is a virtual transition suppressed by an energy denominator of
order
$J$,
so its effect on the low-energy physics is weaker than a retained,
on-resonant (floppy) hop by a relative factor of order
$\varGamma/J$.
The next two subsections make this intuition precise, in two
complementary ways.
First, degenerate perturbation theory shows that
equation~\eqref{eq:supp-approx-H} is exactly the first-order effective
Hamiltonian.
Second, the Schrieffer--Wolff transformation bounds the size of the
terms that first order neglects, confirming that they are indeed the
smaller, off-resonant ones.
Both arguments apply to a general interaction matrix $\bm{J}$ and
longitudinal fields $\bm{h}$.
As solvable examples, we show the projectors of the
(1) one-dimensional chain,
(2) one-dimensional chain with uniform longitudinal fields (PXP model),
(3) two-dimensional square lattice,
(4) $d$-dimensional hypercubic lattice,
and
(5) all-to-all Lipkin--Meshkov--Glick (LMG) model.
For random interaction networks,
we numerically validate the equienergy approximation of the Hamiltonian.

\subsection{Validation of the equienergy approximation using
perturbation theory}
We now derive equation~\eqref{eq:supp-approx-H} from degenerate
perturbation theory.
Split the Hamiltonian as
$\hat{\mathcal{H}}=\hat{\mathcal{H}}_0+\hat{V}$,
with the classical (diagonal) part
\begin{equation}
    \hat{\mathcal{H}}_0
    =
    \sum_{\bm{s}}
    H(\bm{s})
    \ket{\bm{s}}\!\bra{\bm{s}}
\end{equation}
and the transverse field as the perturbation
\begin{align}
    \hat{V}
    &=
    -
    \varGamma
    \sum_{k=1}^{N}
    \hat{X}_k
    \nonumber
    \\
    &=
    -
    \varGamma
    \sum_{\bm{s}}
    \sum_{k=1}^{N}
    \ket{\bm{F}_k \bm{s}}\!\bra{\bm{s}}
    .
\end{align}
The unperturbed eigenstates are the Fock (computational) states
$\ket{\bm{s}}$,
with eigenvalues $H(\bm{s})$.
Because many configurations share the same classical energy,
the spectrum of
$\hat{\mathcal{H}}_0$
is highly degenerate.
Let
\begin{equation}
    \hat{P}_E
    =
    \sum_{\bm{s}:\,H(\bm{s})=E}
    \ket{\bm{s}}\!\bra{\bm{s}}
\end{equation}
be the projector onto the degenerate shell of energy $E$,
and write
$\hat{V}=\hat{V}_{\mathrm{d}}+\hat{V}_{\mathrm{od}}$
with the diagonal (intra-shell) and off-diagonal (inter-shell) parts
\begin{align}
    \hat{V}_{\mathrm{d}}
    &=
    \sum_{E}
    \hat{P}_E
    \hat{V}
    \hat{P}_E
    ,
    \\
    \hat{V}_{\mathrm{od}}
    &=
    \sum_{E\neq E^\prime}
    \hat{P}_E
    \hat{V}
    \hat{P}_{E^\prime}
    .
\end{align}
When
$\hat{\mathcal{H}}_0$
has a degenerate eigenspace,
ordinary (non-degenerate) Rayleigh--Schr\"odinger perturbation theory
breaks down within it:
the usual first-order correction to an eigenstate,
\begin{equation}
    \ket{\bm{s}}^{(1)}
    =
    \sum_{\bm{s}^\prime\neq\bm{s}}
    \frac{\bra{\bm{s}^\prime}\hat{V}\ket{\bm{s}}}{H(\bm{s})-H(\bm{s}^\prime)}
    \ket{\bm{s}^\prime}
    ,
\end{equation}
diverges whenever
$\bm{s}^\prime$
lies in the same shell as
$\bm{s}$,
since the energy denominator vanishes.
The standard remedy is to first diagonalize the perturbation
\emph{within} each degenerate shell:
the correct zeroth-order eigenstates are the eigenvectors of
$\hat{P}_E\hat{V}\hat{P}_E$
restricted to shell
$E$,
and the corresponding first-order energy shifts are eigenvalues of
that operator.
Equivalently---and more convenient here, since it avoids
diagonalizing every shell by hand---one may keep working in the
original Fock basis and simply discard every matrix element of
$\hat{V}$
that connects \emph{different} shells,
i.e., replace
$\hat{V}$
everywhere by
$\hat{V}_{\mathrm{d}}$.
Because
$\hat{V}_{\mathrm{d}}$
has, by its own definition, no matrix elements between shells
($\hat{P}_E\hat{V}_{\mathrm{d}}\hat{P}_{E^\prime}=0$
    for
    $E\neq E^\prime$
),
and
$\hat{\mathcal{H}}_0$
is already diagonal,
the sum
$\hat{\mathcal{H}}_0+\hat{V}_{\mathrm{d}}$
is exactly block-diagonal, one block per shell.
Diagonalizing it block by block therefore reproduces exactly the same
zeroth-order eigenstates and first-order energies that the
shell-by-shell recipe above would give,
without ever having to leave the Fock basis.
This is why,
in each shell,
first-order degenerate perturbation theory reduces to replacing
$\hat{V}$
by its restriction
$\hat{V}_{\mathrm{d}}$.
A matrix element of
$\hat{V}$
connects
$\ket{\bm{F}_k\bm{s}}$
to
$\ket{\bm{s}}$
and therefore contributes to
$\hat{V}_{\mathrm{d}}$
if and only if
$H(\bm{F}_k \bm{s})=H(\bm{s})$,
i.e.,
spin $k$ is floppy in configuration
$\bm{s}$.
Hence
\begin{equation}
    \hat{V}_{\mathrm{d}}
    =
    -
    \varGamma
    \sum_{\bm{s}}
    \sum_{k=1}^{N}
    \delta_{H(\bm{s}), H(\bm{F}_k\bm{s})}
    \ket{\bm{F}_k\bm{s}}\!\bra{\bm{s}}
    ,
\end{equation}
and the first-order effective Hamiltonian in the Fock basis is
precisely equation~\eqref{eq:supp-approx-H},
\begin{equation}
    \hat{\mathcal{H}}_{\mathrm{eff}}^{(1)}
    =
    \hat{\mathcal{H}}_0
    +
    \hat{V}_{\mathrm{d}}
    .
\end{equation}

The second-order correction can be obtained by ordinary
quasi-degenerate perturbation theory,
via L\"owdin--Feshbach partitioning~\cite{Lowdin1962,Feshbach1962},
without invoking a unitary block-diagonalizing transformation.
Again, the standard Rayleigh--Schr\"odinger perturbation theory formula at
second order is ill-defined because of the degeneracy.
Let
$\hat{P}\coloneqq\hat{P}_{H(\bm{s})}$
project onto the shell containing
$\bm{s}$,
and
$\hat{Q}\coloneqq\hat{1}-\hat{P}$
its complement.
Write
$\ket{\psi}=\ket{\psi_P}+\ket{\psi_Q}$
with
$\ket{\psi_P}\coloneqq\hat{P}\ket{\psi}$
and
$\ket{\psi_Q}\coloneqq\hat{Q}\ket{\psi}$.
Projecting the eigenvalue equation
$\hat{\mathcal{H}}\ket{\psi}=E\ket{\psi}$
onto
$\hat{P}$
and onto
$\hat{Q}$,
and using
$\hat{P}\hat{Q}=\hat{Q}\hat{P}=0$,
$\hat{P}^2=\hat{P}$,
$\hat{Q}^2=\hat{Q}$,
$
\hat{P}(\hat{\mathcal{H}}\ket{\psi})
=
\hat{P}\hat{\mathcal{H}}(\hat{P}+\hat{Q})\ket{\psi}
=
\hat{P}\hat{\mathcal{H}}(\hat{P}^2+\hat{Q}^2)\ket{\psi}
=
\hat{P}(E\ket{\psi})
$,
gives the coupled pair
\begin{align*}
    \hat{P}\hat{\mathcal{H}}\hat{P}\ket{\psi_P}
    +
    \hat{P}\hat{\mathcal{H}}\hat{Q}\ket{\psi_Q}
    &=
    E
    \ket{\psi_P}
    ,
    \\
    \hat{Q}\hat{\mathcal{H}}\hat{P}\ket{\psi_P}
    +
    \hat{Q}\hat{\mathcal{H}}\hat{Q}\ket{\psi_Q}
    &=
    E
    \ket{\psi_Q}
    .
\end{align*}
Solving the second line for
$\ket{\psi_Q}$
within the
$\hat{Q}$
subspace gives
\begin{equation*}
    \ket{\psi_Q}
    =
    \frac{1}{E-\hat{Q}\hat{\mathcal{H}}\hat{Q}}
    \hat{Q}\hat{\mathcal{H}}\hat{P}
    \ket{\psi_P}
    ,
\end{equation*}
and substituting into the first line eliminates
$\ket{\psi_Q}$,
leaving a closed eigenvalue equation
$\hat{\mathcal{H}}_{\mathrm{eff}}(E)\ket{\psi_P}=E\ket{\psi_P}$
for
$\ket{\psi_P}$
alone, with
\begin{equation*}
    \hat{\mathcal{H}}_{\mathrm{eff}}(E)
    =
    \hat{P}\hat{\mathcal{H}}\hat{P}
    +
    \hat{P}\hat{\mathcal{H}}\hat{Q}
    \,
    \frac{1}{E-\hat{Q}\hat{\mathcal{H}}\hat{Q}}
    \,
    \hat{Q}\hat{\mathcal{H}}\hat{P}
    .
\end{equation*}
Because
$\hat{\mathcal{H}}_0$
is diagonal in the Fock basis, it is block-diagonal with respect to
$\hat{P}$
and
$\hat{Q}$,
so
$\hat{P}\hat{\mathcal{H}}_0\hat{Q}=0$;
likewise
$\hat{P}\hat{V}_{\mathrm{d}}\hat{Q}=0$,
since
$\hat{V}_{\mathrm{d}}$
has no matrix elements between shells by construction,
and
$\hat{P}\hat{V}_{\mathrm{od}}\hat{P}=0$,
since
$\hat{V}_{\mathrm{od}}$
has none within a shell.
Hence
$\hat{P}\hat{\mathcal{H}}\hat{P}=\hat{P}\hat{\mathcal{H}}_0\hat{P}+\hat{P}\hat{V}_{\mathrm{d}}\hat{P}$
and
$\hat{P}\hat{\mathcal{H}}\hat{Q}=\hat{P}\hat{V}_{\mathrm{od}}\hat{Q}$
(with adjoint
$\hat{Q}\hat{\mathcal{H}}\hat{P}=\hat{Q}\hat{V}_{\mathrm{od}}\hat{P}$),
so
$\hat{\mathcal{H}}_{\mathrm{eff}}(E)$
reduces to the exact effective Hamiltonian
\begin{equation}
    \hat{\mathcal{H}}_{\mathrm{eff}}(E)
    =
    \hat{P}\hat{\mathcal{H}}_0\hat{P}
    +
    \hat{P}\hat{V}_{\mathrm{d}}\hat{P}
    +
    \hat{P}\hat{V}_{\mathrm{od}}\hat{Q}
    \,
    \frac{1}{E-\hat{Q}\hat{\mathcal{H}}\hat{Q}}
    \,
    \hat{Q}\hat{V}_{\mathrm{od}}\hat{P}
    .
\end{equation}
The last term is already
$O(\varGamma^2)$,
so replacing
$E\to H(\bm{s})$
and
$\hat{Q}\hat{\mathcal{H}}\hat{Q}\to\hat{Q}\hat{\mathcal{H}}_0\hat{Q}$
inside it gives, in the Fock basis,
\begin{equation}
    \bra{\bm{s}}
    \hat{\mathcal{H}}_{\mathrm{eff}}^{(2)}
    \ket{\bm{s}^\prime}
    =
    H(\bm{s})
    \delta_{\bm{s}, \bm{s}^\prime}
    +
    \bra{\bm{s}}\hat{V}_{\mathrm{d}}\ket{\bm{s}^\prime}
    +
    \sum_{\bm{s}^{\prime\prime}:\,H(\bm{s}^{\prime\prime})\neq H(\bm{s})}
    \frac{
        \bra{\bm{s}}\hat{V}_{\mathrm{od}}\ket{\bm{s}^{\prime\prime}}
        \bra{\bm{s}^{\prime\prime}}\hat{V}_{\mathrm{od}}\ket{\bm{s}^\prime}
    }{
        H(\bm{s})-H(\bm{s}^{\prime\prime})
    }
\end{equation}
for
$\bm{s},\bm{s}^\prime$
in the same shell.
Because
$H(\bm{s})=H(\bm{s}^\prime)$
there, symmetrizing the denominator,
\begin{equation*}
    \frac{1}{H(\bm{s})-H(\bm{s}^{\prime\prime})}
    \to
    \frac{1}{2}
    \bigg(
        \frac{1}{H(\bm{s})-H(\bm{s}^{\prime\prime})}
        +
        \frac{1}{H(\bm{s}^\prime)-H(\bm{s}^{\prime\prime})}
    \bigg)
    ,
\end{equation*}
makes
$\hat{\mathcal{H}}_{\mathrm{eff}}^{(2)}$
Hermitian within the shell.
Every term in the sum has
$H(\bm{s}^{\prime\prime})\neq H(\bm{s})$
by construction,
since
$\hat{V}_{\mathrm{od}}$
excludes same-shell connections;
the leading correction is therefore well-defined without further
degenerate treatment,
and the Schrieffer--Wolff calculation below reproduces this same
$\hat{\mathcal{H}}_{\mathrm{eff}}^{(2)}$
order by order in
$\varGamma/J$.

\subsection{Validation of the equienergy approximation using the
Schrieffer--Wolff transformation}
The Schrieffer--Wolff (SW) transformation~\cite{Schrieffer1966}
provides a systematic unitary that block-diagonalizes
$\hat{\mathcal{H}}$
with respect to the classical energy shells,
order by order in
$\varGamma/J$.
One seeks an anti-Hermitian generator
$\hat{S}=-\hat{S}^\dagger$
such that the Baker--Campbell--Hausdorff expansion
\begin{equation}
    \exp(\hat{S})
    \hat{\mathcal{H}}
    \exp(-\hat{S})
    =
    \hat{\mathcal{H}}_0
    +
    \hat{V}
    +
    [
        \hat{S}
        ,
        \hat{\mathcal{H}}_0
        +
        \hat{V}
    ]
    +
    \frac{1}{2}
    [
        \hat{S}
        ,
        [
            \hat{S}
            ,
            \hat{\mathcal{H}}_0
            +
            \hat{V}
        ]
    ]
    +
    \cdots
\end{equation}
has vanishing off-diagonal blocks up to the desired order.
At leading order one chooses
$\hat{S}$
to cancel
$\hat{V}_{\mathrm{od}}$,
namely
\begin{equation}
    [
        \hat{S}
        ,
        \hat{\mathcal{H}}_0
    ]
    +
    \hat{V}_{\mathrm{od}}
    =
    0
    .
\end{equation}
In the Fock basis this fixes
\begin{equation}
    \bra{\bm{s}}
    \hat{S}
    \ket{\bm{s}^\prime}
    =
    \frac{
        \bra{\bm{s}}
        \hat{V}_{\mathrm{od}}
        \ket{\bm{s}^\prime}
    }{
        H(\bm{s}^\prime)
        -
        H(\bm{s})
    }
\end{equation}
whenever
$H(\bm{s})\neq H(\bm{s}^\prime)$,
and zero otherwise.
The second-order effective Hamiltonian inside each shell then reads
\begin{equation}
    \hat{\mathcal{H}}_{\mathrm{eff}}^{(2)}
    =
    \hat{\mathcal{H}}_0
    +
    \hat{V}_{\mathrm{d}}
    +
    \frac{1}{2}
    \hat{P}
    [
        \hat{S}
        ,
        \hat{V}_{\mathrm{od}}
    ]
    \hat{P}
    ,
\end{equation}
where
$\hat{P}$
denotes the projector onto the shell of interest.
For a virtual transition
$\bm{s}\to\bm{s}^{\prime\prime}\to\bm{s}^\prime$
with
$H(\bm{s})=H(\bm{s}^\prime)\neq H(\bm{s}^{\prime\prime})$,
the induced amplitude is of order
\begin{equation}
    \frac{
        \varGamma^2
    }{
        H(\bm{s})
        -
        H(\bm{s}^{\prime\prime})
    }
    .
\end{equation}
Under the weak-field condition
$|\varGamma|\ll |J|$,
a typical off-resonant denominator is
$|H(\bm{F}_k \bm{s})-H(\bm{s})|\sim |J|$,
so these corrections are
$O(\varGamma^2/J)$
and are small compared with the resonant
$O(\varGamma)$
hopping retained in
$\hat{V}_{\mathrm{d}}$.

For the discrete couplings of the main text,
$\bm{J}\in\{0,+J\}^{N\times N}$
and
$\bm{h}\in\{+J,+2J\}^N$,
the classical energy differences are integer multiples of $J$.
Flipping spin $k$ changes the energy by
\begin{equation}
    H(\bm{F}_k\bm{s})
    -
    H(\bm{s})
    =
    2
    s_k
    \Bigg(
        \sum_{j=1}^{N}
        J_{k,j}
        s_j
        +
        h_k
    \Bigg)
    ,
\end{equation}
so every non-floppy flip has
\begin{equation}
    |
    H(\bm{F}_k\bm{s})
    -
    H(\bm{s})
    |
    \geq
    2J
    .
\end{equation}
The SW generator is therefore bounded by
$|\varGamma|/2J$,
and the second-order corrections are at most
$O(\varGamma^2/2J)$.
With the working value
$\varGamma=0.1J$
this is
$O(5\times 10^{-3}\,J)$,
an order of magnitude below the retained floppy hopping
$\varGamma=0.1J$.
Truncating at first order in
$\varGamma/J$
therefore recovers equation~\eqref{eq:supp-approx-H}
with a controlled error for the discrete ensemble used throughout
the paper.

\subsection{Application to the transverse-field Ising spin systems}
We then apply the equienergy approximation to the transverse-field
Ising spin systems to show that it reproduces the known projectors.
Below we simply write the periodic boundary conditions without using
the $\bmod$ operator for clarity, but readers should remember that
the system is periodic, e.g., $i+1$ should be interpreted as $(i+1)\bmod N$.
\subsubsection{Transverse-field Ising chain}
We apply the above reasoning to the chain
\begin{equation}
    \hat{\mathcal{H}}
    =
    -
    J
    \sum_{i=1}^{N}
    \hat{Z}_i
    \hat{Z}_{i+1}
    -
    \varGamma
    \sum_{i=1}^{N}
    \hat{X}_i
    .
\end{equation}
The classical energy of a configuration
$\bm{s}\in\{+1,-1\}^N$
is
\begin{equation}
    H(\bm{s})
    =
    -
    J
    \sum_{i=1}^{N}
    s_i
    s_{i+1}
    .
\end{equation}
Flipping spin $k$ changes the energy by
\begin{equation}
    H(\bm{F}_k\bm{s})
    -
    H(\bm{s})
    =
    2
    J
    s_k
    (
        s_{k-1}
        +
        s_{k+1}
    )
    .
\end{equation}
Hence, the floppy condition
$H(\bm{F}_k\bm{s})=H(\bm{s})$
holds if and only if
\begin{equation}
    s_{k-1}
    +
    s_{k+1}
    =
    0
    ,
\end{equation}
i.e.,
the two neighbours of site $k$ point in opposite directions.
Such a site sits next to a domain wall:
flipping it moves the domain wall by one lattice spacing and leaves $H(\bm{s})$
unchanged.
The indicator of the floppy condition is therefore
\begin{align}
    \delta_{H(\bm{s}), H(\bm{F}_k\bm{s})}
    &=
    \delta_{0, s_{k-1}+s_{k+1}}
    \\
    &=
    1
    -
    \bigg(
        \frac{s_{k-1}+s_{k+1}}{2}
    \bigg)^2
    \\
    &=
    \frac{1-s_{k-1}s_{k+1}}{2}
    .
\end{align}
Note that $s_k^2=1$.
We define the projector
\begin{align}
    \hat{P}_i
    &\coloneqq
    \frac{\hat{1}-\hat{Z}_{i-1}\hat{Z}_{i+1}}{2}
    \\
    &=
    \ket{\uparrow_{i-1}\downarrow_{i+1}}\!\bra{\uparrow_{i-1}\downarrow_{i+1}}
    +
    \ket{\downarrow_{i-1}\uparrow_{i+1}}\!\bra{\downarrow_{i-1}\uparrow_{i+1}}
    ,
\end{align}
which satisfies
$\hat{P}_i\ket{\bm{s}}=[(1-s_{i-1}s_{i+1})/2]\ket{\bm{s}}$.
The flip $\bm{F}_k$ changes only $s_k$,
so $\hat{P}_k$,
which acts only on sites $k\pm1$,
commutes with $\hat{X}_k$ and takes the same value on
$\ket{\bm{s}}$ and $\ket{\bm{F}_k\bm{s}}$.
Using
$\sum_{\bm{s}}\ket{\bm{F}_k \bm{s}}\!\bra{\bm{s}}=\hat{X}_k$,
we have
\begin{align}
    \sum_{\bm{s}}
    \delta_{H(\bm{s}), H(\bm{F}_k\bm{s})}
    \ket{\bm{F}_k \bm{s}}\!\bra{\bm{s}}
    &=
    \sum_{\bm{s}}
    \frac{\hat{1}-\hat{Z}_{k-1}\hat{Z}_{k+1}}{2}
    \ket{\bm{F}_k \bm{s}}\!\bra{\bm{s}}
    \nonumber
    \\
    &=
    \frac{1}{2}
    \hat{X}_k
    (
        \hat{1}
        -
        \hat{Z}_{k-1}
        \hat{Z}_{k+1}
    )
    .
\end{align}
The intra-shell part of the perturbation is then
\begin{equation}
    \hat{V}_{\mathrm{d}}
    =
    -
    \varGamma
    \frac{1}{2}
    \sum_{k=1}^{N}
    \hat{X}_k
    (
        \hat{1}
        -
        \hat{Z}_{k-1}
        \hat{Z}_{k+1}
    )
    ,
\end{equation}
as derived in refs.~\cite{Lee2005,Yoshinaga2022},
and can be mapped~\cite{Ostmann2019} to the $XX$ model (or free fermions)
\begin{equation}
    \hat{\mathcal{H}}_{XX}
    \coloneqq
    \frac{\varGamma}{2}
    \sum_{k=1}^{N}
    (
        \hat{X}_k
        \hat{X}_{k+1}
        +
        \hat{Y}_k
        \hat{Y}_{k+1}
    )
\end{equation}
through the Kramers--Wannier transformation.

This picture is consistent with the exact Bogoliubov spectrum.
Indeed, domain walls in this real-space, floppy-hop picture
are the Bogoliubov quasiparticles of that momentum-space calculation:
both are the elementary excitations created above the ferromagnetic
ground state,
just described in two different bases.
For
$|\varGamma|\ll J$,
one expands
\begin{align}
    E_\kappa
    &=
    2
    \sqrt{
        J^2
        +
        \varGamma^2
        -
        2
        J
        \varGamma
        \cos(\kappa)
    }
    \nonumber
    \\
    &=
    2
    J
    \sqrt{
        1
        -
        2
        \frac{\varGamma}{J}
        \cos(\kappa)
        +
        \bigg(
            \frac{\varGamma}{J}
        \bigg)^2
    }
    \nonumber
    \\
    &=
    2
    J
    -
    2
    \varGamma
    \cos(\kappa)
    +
    O\bigg(
        \frac{\varGamma^2}{J}
    \bigg)
    .
\end{align}
The leading correction
$-2 \varGamma\cos(\kappa)$
is the dispersion of a tight-binding domain wall with nearest-neighbour
hopping
$-\varGamma$,
while the remainder is of the same order
$O(\varGamma^2/J)$
as the SW corrections discarded by the projection.
Likewise,
setting
$\kappa=0$
in the exact dispersion gives
$E_{\kappa=0}=2\sqrt{J^2+\varGamma^2-2J\varGamma}=2\sqrt{(J-\varGamma)^2}=2|J-\varGamma|$,
which reduces to
$2J+O(\varGamma)$
on the ferromagnetic side,
matching the classical cost of inserting domain walls up to weak-field
shifts.

\subsubsection{PXP model: Weak-transverse-field Ising chain with
longitudinal field}
Following refs.~\cite{Yoshinaga2022,Peng2022a},
we consider the weak-transverse-field Ising chain with the longitudinal field,
$h_i=2J$ for all $i$,
\begin{equation}
    \hat{\mathcal{H}}
    =
    -
    J
    \sum_{i=1}^{N}
    \hat{Z}_i
    \hat{Z}_{i+1}
    -
    2J
    \sum_{i=1}^{N}
    \hat{Z}_i
    -
    \varGamma
    \sum_{i=1}^{N}
    \hat{X}_i
    .
\end{equation}
The corresponding classical energy is
\begin{equation}
    H(\bm{s})
    =
    -
    J
    \sum_{i=1}^{N}
    s_i
    s_{i+1}
    -
    2J
    \sum_{i=1}^{N}
    s_i
    .
\end{equation}
Flipping spin $k$ changes the energy by
\begin{align}
    H(\bm{F}_k\bm{s})
    -
    H(\bm{s})
    &=
    2J
    s_k
    (
        s_{k-1}
        +
        s_{k+1}
    )
    +
    4J
    s_k
    \nonumber
    \\
    &=
    2J
    s_k
    (
        s_{k-1}
        +
        s_{k+1}
        +
        2
    )
    .
\end{align}
The floppy condition is
\begin{equation}
    s_{k-1}
    +
    s_{k+1}
    +
    2
    =
    0
    ,
\end{equation}
that is,
the two neighbours of site $k$ are both in the down state.
Since $s_{k\pm1}\in\{+1,-1\}$,
the sum equals $-2$ only if $s_{k-1}=s_{k+1}=-1$.
Hence, the indicator of the floppy condition is
\begin{align}
    \delta_{H(\bm{s}), H(\bm{F}_k\bm{s})}
    &=
    \delta_{0, s_{k-1}+s_{k+1}+2}
    \nonumber
    \\
    &=
    \delta_{-1,s_{k-1}}
    \delta_{-1,s_{k+1}}
    \nonumber
    \\
    &=
    \frac{1-s_{k-1}}{2}
    \frac{1-s_{k+1}}{2}
    .
\end{align}
We define the projector onto the down state of site $i$,
\begin{align}
    \hat{P}_i
    &\coloneqq
    \frac{\hat{1}-\hat{Z}_i}{2}
    \\
    &=
    \ket{\downarrow_i}\!\bra{\downarrow_i}
    ,
\end{align}
which satisfies
$\hat{P}_i\ket{\bm{s}}=[(1-s_i)/2]\ket{\bm{s}}$.
The flip $\bm{F}_k$ changes only $s_k$,
so $\hat{P}_{k\pm1}$ commutes with $\hat{X}_k$ and takes the same value on
$\ket{\bm{s}}$ and $\ket{\bm{F}_k\bm{s}}$.
Using
$\sum_{\bm{s}}\ket{\bm{F}_k \bm{s}}\!\bra{\bm{s}}=\hat{X}_k$,
\begin{align}
    \sum_{\bm{s}}
    \delta_{H(\bm{s}), H(\bm{F}_k\bm{s})}
    \ket{\bm{F}_k \bm{s}}\!\bra{\bm{s}}
    &=
    \sum_{\bm{s}}
    \hat{P}_{k-1}
    \ket{\bm{F}_k \bm{s}}\!\bra{\bm{s}}
    \hat{P}_{k+1}
    \nonumber
    \\
    &=
    \hat{P}_{k-1}
    \hat{X}_k
    \hat{P}_{k+1}
    .
\end{align}
Substituting into the definition of $\hat{V}_{\mathrm{d}}$
gives the PXP Hamiltonian
\begin{equation}
    \hat{V}_{\mathrm{d}}
    =
    -
    \varGamma
    \sum_{k=1}^{N}
    \hat{P}_{k-1}
    \hat{X}_k
    \hat{P}_{k+1}
    ,
\end{equation}
as derived in ref.~\cite{Yoshinaga2022}.
Because
$\hat{V}_{\mathrm{d}}$
connects only configurations of equal classical energy,
$[\hat{\mathcal{H}}_0,\hat{V}_{\mathrm{d}}]=0$,
and inside each energy shell the effective Hamiltonian
$\hat{\mathcal{H}}_0+\hat{V}_{\mathrm{d}}$
reduces to $\hat{V}_{\mathrm{d}}$ up to the constant $H(\bm{s})$.
A spin can be flipped from down to up only if both neighbours are down,
so no two adjacent up spins are created.
The sector without adjacent up spins is therefore closed under
$\hat{V}_{\mathrm{d}}$,
and there $\hat{V}_{\mathrm{d}}$ is the Rydberg-blockade PXP
model~\cite{Bernien2017,Turner2018,Turner2018a}
with the up state corresponding to the excited state.

\subsubsection{Transverse-field Ising model on a two-dimensional lattice}
The Hamiltonian of the transverse-field Ising model on an $L\times L$
square lattice with periodic boundary conditions is
\begin{equation}
    \hat{\mathcal{H}}
    =
    -
    J
    \sum_{i=1}^{L^2}
    \hat{Z}_{i}
    (
        \hat{Z}_{i+1}
        +
        \hat{Z}_{i+L}
    )
    -
    \varGamma
    \sum_{i=1}^{L^2}
    \hat{X}_i
    .
\end{equation}
The classical energy of a configuration
$\bm{s}\in\{+1,-1\}^{L^2}$
is
\begin{equation}
    H(\bm{s})
    =
    -
    J
    \sum_{i=1}^{L^2}
    s_{i}
    (
        s_{i+1}
        +
        s_{i+L}
    )
    .
\end{equation}
Flipping spin $k$ changes the energy by
\begin{equation}
    H(\bm{F}_k\bm{s})
    -
    H(\bm{s})
    =
    2
    J
    s_k
    (
        s_{k-1}
        +
        s_{k+1}
        +
        s_{k-L}
        +
        s_{k+L}
    )
    .
\end{equation}
The floppy condition gives
\begin{equation}
    s_{k-1}
    +
    s_{k+1}
    +
    s_{k-L}
    +
    s_{k+L}
    =
    0
    ,
\end{equation}
which means that the sum of the spins of the four nearest neighbours
of spin $k$ is zero.
Each spin takes the values $\pm1$,
so the sum of the four neighbouring spins takes only the values
$-4$, $-2$, $0$, $2$, and $4$.
The polynomial below vanishes at $\pm2$ and $\pm4$ and equals $1$ at $0$,
so the indicator of the floppy condition is
\begin{align}
    \delta_{H(\bm{s}), H(\bm{F}_k\bm{s})}
    &=
    \delta_{0, s_{k-1}+s_{k+1}+s_{k-L}+s_{k+L}}
    \\
    &=
    \bigg[
        1
        -
        \bigg(
            \frac{s_{k-1}+s_{k+1}+s_{k-L}+s_{k+L}}{2}
        \bigg)^2
    \bigg]
    \bigg[
        1
        -
        \bigg(
            \frac{s_{k-1}+s_{k+1}+s_{k-L}+s_{k+L}}{4}
        \bigg)^2
    \bigg]
    \\
    &=
    \frac{1}{64}
    [
        4
        -
        (
            s_{k-1}
            +
            s_{k+1}
            +
            s_{k-L}
            +
            s_{k+L}
        )^2
    ]
    [
        16
        -
        (
            s_{k-1}
            +
            s_{k+1}
            +
            s_{k-L}
            +
            s_{k+L}
        )^2
    ]
    \nonumber
    \\
    &=
    \frac{1}{64}
    [
        64
        -
        20
        (
            s_{k-1}
            +
            s_{k+1}
            +
            s_{k-L}
            +
            s_{k+L}
        )^2
        +
        (
            s_{k-1}
            +
            s_{k+1}
            +
            s_{k-L}
            +
            s_{k+L}
        )^4
    ]
    .
    \label{eq:2d-indicator}
\end{align}
We expand equation~\eqref{eq:2d-indicator} using $s_i^2=1$.
The square of the neighbour sum is
\begin{equation}
    (
        s_{k-1}
        +
        s_{k+1}
        +
        s_{k-L}
        +
        s_{k+L}
    )^2
    =
    4
    +
    2
    (
        s_{k-1}s_{k+1}
        +
        s_{k-1}s_{k-L}
        +
        s_{k-1}s_{k+L}
        +
        s_{k+1}s_{k-L}
        +
        s_{k+1}s_{k+L}
        +
        s_{k-L}s_{k+L}
    )
    \label{eq:2d-sum-squared}
    .
\end{equation}
The fourth power of the neighbour sum is
\begin{align}
    (
        s_{k-1}
        +
        s_{k+1}
        +
        s_{k-L}
        +
        s_{k+L}
    )^4
    &=
    [
        4
        +
        2
        (
            s_{k-1}s_{k+1}
            +
            s_{k-1}s_{k-L}
            +
            s_{k-1}s_{k+L}
            +
            s_{k+1}s_{k-L}
            +
            s_{k+1}s_{k+L}
            +
            s_{k-L}s_{k+L}
        )
    ]^2
    \nonumber
    \\
    &=
    16
    +
    16
    (
        s_{k-1}s_{k+1}
        +
        s_{k-1}s_{k-L}
        +
        s_{k-1}s_{k+L}
        +
        s_{k+1}s_{k-L}
        +
        s_{k+1}s_{k+L}
        +
        s_{k-L}s_{k+L}
    )
    \nonumber
    \\&\qquad~
    +
    4
    (
        s_{k-1}s_{k+1}
        +
        s_{k-1}s_{k-L}
        +
        s_{k-1}s_{k+L}
        +
        s_{k+1}s_{k-L}
        +
        s_{k+1}s_{k+L}
        +
        s_{k-L}s_{k+L}
    )^2
    \nonumber
    \\
    &=
    16
    +
    4
    \cdot
    6
    +
    16
    (
        s_{k-1}s_{k+1}
        +
        s_{k-1}s_{k-L}
        +
        s_{k-1}s_{k+L}
        +
        s_{k+1}s_{k-L}
        +
        s_{k+1}s_{k+L}
        +
        s_{k-L}s_{k+L}
    )
    \nonumber
    \\&\qquad~
    +
    8
    s_{k-1}s_{k+1}
    (
        s_{k-1}s_{k-L}
        +
        s_{k-1}s_{k+L}
        +
        s_{k+1}s_{k-L}
        +
        s_{k+1}s_{k+L}
        +
        s_{k-L}s_{k+L}
    )
    \nonumber
    \\&\qquad~
    +
    8
    s_{k-1}s_{k-L}
    (
        s_{k-1}s_{k+L}
        +
        s_{k+1}s_{k-L}
        +
        s_{k+1}s_{k+L}
        +
        s_{k-L}s_{k+L}
    )
    \nonumber
    \\&\qquad~
    +
    8
    s_{k-1}s_{k+L}
    (
        s_{k+1}s_{k-L}
        +
        s_{k+1}s_{k+L}
        +
        s_{k-L}s_{k+L}
    )
    \nonumber
    \\&\qquad~
    +
    8
    s_{k+1}s_{k-L}
    (
        s_{k+1}s_{k+L}
        +
        s_{k-L}s_{k+L}
    )
    \nonumber
    \\&\qquad~
    +
    8
    s_{k+1}s_{k+L}
    s_{k-L}s_{k+L}
    \nonumber
    \\
    &=
    40
    +
    16
    (
        s_{k-1}s_{k+1}
        +
        s_{k-1}s_{k-L}
        +
        s_{k-1}s_{k+L}
        +
        s_{k+1}s_{k-L}
        +
        s_{k+1}s_{k+L}
        +
        s_{k-L}s_{k+L}
    )
    \nonumber
    \\&\qquad~
    +
    8
    (
        s_{k+1} s_{k-L}
        +
        s_{k+1} s_{k+L}
        +
        s_{k-1} s_{k-L}
        +
        s_{k-1} s_{k+L}
        +
        s_{k-1}s_{k+1} s_{k-L}s_{k+L}
    )
    \nonumber
    \\&\qquad~
    +
    8
    (
        s_{k-L} s_{k+L}
        +
        s_{k-1} s_{k+1}
        +
        s_{k-1}s_{k+1}s_{k-L}s_{k+L}
        +
        s_{k-1} s_{k+L}
    )
    \nonumber
    \\&\qquad~
    +
    8
    (
        s_{k-1}s_{k+1}s_{k-L}s_{k+L}
        +
        s_{k-1} s_{k+1}
        +
        s_{k-1} s_{k-L}
    )
    \nonumber
    \\&\qquad~
    +
    8
    (
        s_{k-L} s_{k+L}
        +
        s_{k+1} s_{k+L}
    )
    \nonumber
    \\&\qquad~
    +
    8
    s_{k+1} s_{k-L}
    \nonumber
    \\
    &=
    40
    +
    32
    (
        s_{k-1}s_{k+1}
        +
        s_{k-1}s_{k-L}
        +
        s_{k-1}s_{k+L}
        +
        s_{k+1}s_{k-L}
        +
        s_{k+1}s_{k+L}
        +
        s_{k-L}s_{k+L}
    )
    \nonumber
    \\&\qquad~
    +
    24
    s_{k-1}s_{k+1} s_{k-L}s_{k+L}
    .
    \label{eq:2d-sum-fourth}
\end{align}
Substituting these two results
[equations~\eqref{eq:2d-sum-squared} and~\eqref{eq:2d-sum-fourth}]
into equation~\eqref{eq:2d-indicator} gives
\begin{align}
    \delta_{H(\bm{s}), H(\bm{F}_k\bm{s})}
    &=
    \frac{1}{8}
    [
        3
        -
        (
            s_{k-1}s_{k+1}
            +
            s_{k-1}s_{k-L}
            +
            s_{k-1}s_{k+L}
            +
            s_{k+1}s_{k-L}
            +
            s_{k+1}s_{k+L}
            +
            s_{k-L}s_{k+L}
        )
        +
        3
        s_{k-1}
        s_{k+1}
        s_{k-L}
        s_{k+L}
    ]
    \\
    &=
    \frac{5}{8}
    -
    \frac{1}{16}
    (
        s_{k-1}
        +
        s_{k+1}
        +
        s_{k-L}
        +
        s_{k+L}
    )^2
    +
    \frac{3}{8}
    s_{k-1}
    s_{k+1}
    s_{k-L}
    s_{k+L}
    .
\end{align}
Replacing each $s_i$ by $\hat{Z}_i$, we obtain the projector
\begin{align}
    \hat{P}_k
    \coloneqq
    \frac{5}{8}
    -
    \frac{1}{16}
    (
        \hat{Z}_{k-1}
        +
        \hat{Z}_{k+1}
        +
        \hat{Z}_{k-L}
        +
        \hat{Z}_{k+L}
    )^2
    +
    \frac{3}{8}
    \hat{Z}_{k-1}
    \hat{Z}_{k+1}
    \hat{Z}_{k-L}
    \hat{Z}_{k+L}
    ,
\end{align}
as derived in ref.~\cite{Yoshinaga2022}.
The flip $\bm{F}_k$ changes only $s_k$,
so $\hat{P}_k$,
which acts only on the neighbours of site $k$,
commutes with $\hat{X}_k$ and takes the same value on
$\ket{\bm{s}}$ and $\ket{\bm{F}_k\bm{s}}$.
Using
$\sum_{\bm{s}}\ket{\bm{F}_k \bm{s}}\!\bra{\bm{s}}=\hat{X}_k$,
\begin{equation}
    \sum_{\bm{s}}
    \delta_{H(\bm{s}), H(\bm{F}_k\bm{s})}
    \ket{\bm{F}_k \bm{s}}\!\bra{\bm{s}}
    =
    \hat{P}_k
    \hat{X}_k
    ,
\end{equation}
and the intra-shell part of the perturbation is
\begin{equation}
    \hat{V}_{\mathrm{d}}
    =
    -
    \varGamma
    \sum_{k=1}^{L^2}
    \hat{P}_k
    \hat{X}_k
    .
\end{equation}
A spin with two up and two down neighbours has two satisfied and two
unsatisfied bonds,
and flipping it exchanges the two.
Hence,
$\hat{V}_{\mathrm{d}}$ deforms domain walls while preserving their total
length,
which is the number of unsatisfied bonds and fixes $H(\bm{s})$.

\subsubsection{Transverse-field Ising model on a $d$-dimensional lattice}
The Hamiltonian of the transverse-field Ising model on a $d$-dimensional
hypercubic lattice of $L^d$ sites with periodic boundary conditions is
\begin{align}
    \hat{\mathcal{H}}
    =
    -
    J
    \sum_{i=1}^{L^d}
    \sum_{\ell=1}^{d}
    \hat{Z}_{i}
    \hat{Z}_{i+L^{\ell-1}}
    -
    \varGamma
    \sum_{i=1}^{L^d}
    \hat{X}_i
    ,
\end{align}
where $i\pm L^{\ell-1}$ is the nearest neighbour of site $i$ along the
$\ell$th axis.
We denote the set of the $2d$ nearest neighbours of site $k$,
which are distinct for $L\geq3$,
by $\mathcal{N}_k$ and their spin sum by
\begin{align}
    \varSigma_k
    \coloneqq
    \sum_{j\in\mathcal{N}_k}
    s_j
    =
    \sum_{\ell=1}^{d}
    (
        s_{k-L^{\ell-1}}
        +
        s_{k+L^{\ell-1}}
    )
    .
\end{align}
Flipping spin $k$ changes the energy by
$H(\bm{F}_k\bm{s})-H(\bm{s})=2J\varSigma_k s_k$,
so the floppy condition is $\varSigma_k=0$.
The number of neighbours $2d$ is even,
so $\varSigma_k=2(N_{\uparrow}-d)$ takes only the values
$0,\pm2,\pm4,\ldots,\pm2d$,
where $N_{\uparrow}$ is the number of up spins in $\mathcal{N}_k$.
The polynomial below vanishes at $\pm2n$ ($n=1,\ldots,d$) and equals $1$
at $0$,
so the indicator of the floppy condition is
\begin{align}
    \delta_{H(\bm{s}), H(\bm{F}_k\bm{s})}
    =
    \prod_{n=1}^{d}
    \bigg[
        1
        -
        \bigg(
            \frac{\varSigma_k}{2n}
        \bigg)^2
    \bigg]
    ,
    \label{eq:dd-indicator}
\end{align}
Replacing each $s_i$ by $\hat{Z}_i$, we obtain the projector
\begin{align}
    \hat{P}_k
    \coloneqq
    \prod_{n=1}^{d}
    \bigg[
        1
        -
        \bigg(
            \frac{\hat{\varSigma}_k}{2n}
        \bigg)^2
    \bigg]
    ,
    \quad
    \hat{\varSigma}_k
    \coloneqq
    \sum_{j\in\mathcal{N}_k}
    \hat{Z}_j
    ,
    \label{eq:dd-projector}
\end{align}
in agreement with ref.~\cite{Yoshinaga2022}.

\subsubsection{The Lipkin--Meshkov--Glick model}
We apply the equienergy approximation to the Lipkin--Meshkov--Glick (LMG)
model~\cite{Lipkin1965}
\begin{equation}
    \hat{\mathcal{H}}
    =
    -
    \frac{1}{2}
    \frac{J}{N}
    \sum_{i=1}^{N}
    \sum_{j=1}^{N}
    \hat{Z}_i
    \hat{Z}_j
    -
    \varGamma
    \sum_{i=1}^{N}
    \hat{X}_i
    .
\end{equation}
The classical energy of a Fock configuration depends only on the
normalized magnetization
$m=(1/N)\sum_{i=1}^{N}s_i\in\{-1, -1+2/N, -1+4/N, \ldots,1\}$,
\begin{equation}
    H(\bm{s})
    =
    -
    N
    \frac{J}{2}
    m^2
    .
\end{equation}
Flipping spin $k$ changes the normalized magnetization by
$-2s_k/N$
and the energy by
\begin{equation}
    H(\bm{F}_k\bm{s})
    -
    H(\bm{s})
    =
    2J
    \bigg(
        m
        s_k
        -
        \frac{1}{N}
    \bigg)
    ,
\end{equation}
so the resonant (floppy) condition
$H(\bm{F}_k\bm{s})=H(\bm{s})$
is
\begin{equation}
    m
    =
    \frac{s_k}{N}
    .
\end{equation}
To see when this can hold, write
$
Nm
=
\sum_{i=1}^N s_i
=
N_\uparrow-N_\downarrow
=
N-2N_\downarrow
$,
where
$N_\uparrow$ and
$N_\downarrow$
are the number of up and down spins, respectively.
Since
$2N_\downarrow$
is always even,
$Nm$
has the same parity as
$N$,
$Nm\equiv N\pmod{2}$.
The floppy condition
$m=s_k/N$
requires
$Nm=\pm1$,
which is odd, so it can be satisfied only when
$N$
itself is odd,
and only inside the classical shell
$|m|=1/N$:
flipping an up spin at
$m=+1/N$
(or a down spin at
$m=-1/N$)
exchanges the two magnetizations while leaving
$H(\bm{s})$
unchanged.
The first-order effective Hamiltonian therefore reduces to
\begin{equation}
    \hat{\mathcal{H}}_{\mathrm{eff}}^{(1)}
    =
    \sum_{\bm{s}}
    H(\bm{s})
    \ket{\bm{s}}\!\bra{\bm{s}}
    -
    \varGamma
    \sum_{\bm{s}}
    \sum_{k=1}^{N}
    \delta_{m, s_k/N}
    \ket{\bm{F}_k \bm{s}}\!\bra{\bm{s}}
    ,
\end{equation}
again of the projected form [equation~\eqref{eq:supp-approx-H}].
For even $N$ no single spin is floppy, so
$\hat{V}_{\mathrm{d}}=0$
and the leading transverse-field part is deferred to the
$O(\varGamma^2)$
Schrieffer--Wolff terms---consistent with the general counting.

\subsubsection{Numerical validation on the random interaction networks}
\begin{figure*}[tb]
    \centering
    \includegraphics{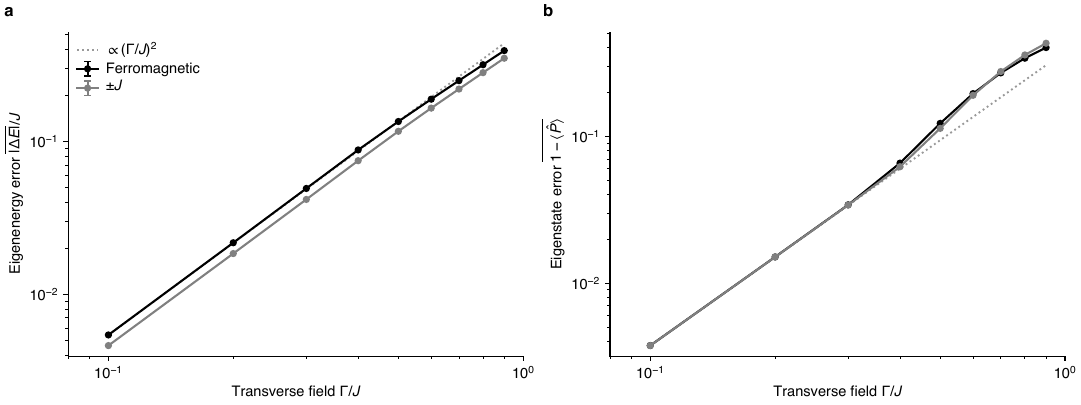}
    \bcaption{
        Validation of the equienergy approximation of the Hamiltonian,
        equation~\eqref{eq:supp-approx-H}.
    }{
        \textbf{a},
        Eigenvalue error
        $
        \overline{|\varDelta E|}/J
        =
        [1/\dim(\mathrm{Hilbert})]
        \sum_i |E_{\mathrm{exact},i}
        -
        E_{\mathrm{approx},i}|/J
        $,
        exact and approximate spectra rank-paired ascending.
        \textbf{b},
        Eigenvector error
        $
        \overline{1 - \langle \hat{P} \rangle}
        =
        [1/\dim(\mathrm{Hilbert})]
        \sum_i
        \bigl(1 - \braket{\psi_{\mathrm{exact},i}
        | \hat{P}_i | \psi_{\mathrm{exact},i}}\bigr)
        $,
        same pairing,
        where $\hat{P}_i$ projects onto the Fock states with the classical
        energy of the $i$th approximate eigenstate.
        Unlike the overlap
        $|\braket{\psi_{\mathrm{exact},i} | \psi_{\mathrm{approx},i}}|^2$,
        it does not depend on the basis chosen within degenerate levels.
        For random Erd\H{o}s--R\'enyi networks at the
        target $\phi = 1$, $N = 12$ ($\dim(\mathrm{Hilbert}) = 4096$),
        $10^4$ independent realizations per point,
        for both ferromagnetic and $\pm J$ coupling ensembles.
        Error bars, standard error of the mean
        (smaller than the markers almost everywhere).
        Dotted line, a $(\varGamma/J)^2$ reference anchored to the
        ferromagnetic ensemble's smallest-$\varGamma$ point in each panel,
        the scaling
        predicted by second-order degenerate perturbation theory
        (Sec.~\ref{sec:validation-approx-h}).
    }
    \label{fig:fig-s01}
\end{figure*}

It is difficult to treat networks with random interactions analytically,
so we numerically validate the equienergy approximation of the Hamiltonian
[equation~\eqref{eq:supp-approx-H}].
In Extended Data Fig.~\ref{fig:fig-s01}, we compare the exact eigenenergies and
eigenstates of the full Hamiltonian with those of the approximated one.
As expected, the eigenenergies and eigenstates of the approximated
Hamiltonian are in good agreement with those of the full Hamiltonian
for small $\varGamma$,
and the error is proportional to $\varGamma^2$, as predicted by the
Schrieffer--Wolff transformation.

\section{Projecting hypercubes}
For the projection of the hypercube, i.e., the Fock-space landscape in
Fig.~\ref{fig:fig-01}f in the main text,
we use the method introduced in ref.~\cite{Horiike2025}.
We perform the weighted principal component analysis (PCA) of the
$N$-dimensional hypercubic vertices $\bm{s} \in \{+1, -1\}^N$.
In the PCA, we first calculate the covariance matrix
\begin{equation}
    \bm{\varSigma}
    =
    \langle
    (
        \bm{s} - \langle\bm{s}\rangle
    )
    (
        \bm{s} - \langle\bm{s}\rangle
    )^\top
    \rangle
    .
\end{equation}
Here, $\langle\cdot\rangle\coloneqq \sum_{\bm{s}} w(\bm{s}) \cdot$
is the weighted average over all hypercubic vertices,
where $w(\bm{s}) \in \mathbb{R}$ is the weight of
the vertex $\bm{s}$, satisfying $0\leq w(\bm{s}) \leq 1$ and
$\sum_{\bm{s}} w(\bm{s}) = 1$.
We weight each vertex by its number of floppy spins,
\begin{equation}
    w(\bm{s})
    \propto
    N f(\bm{s})
    +
    |\xi(\bm{s})|
    ,
\end{equation}
so that the projection resolves the kinetically accessible subspaces.
Here,
$f(\bm{s})$ is the floppiness [equation~\eqref{eq:floppiness} in
the main text]
of the vertex $\bm{s}$,
and $\xi(\bm{s})$ is a Gaussian random number with unit variance and mean
$2^{-N} \sum_{\bm{s}} N f(\bm{s})$,
which lifts the degeneracy of the weights.
We then perform the eigenvalue decomposition
\begin{equation}
    \bm{\varSigma}
    =
    \bm{U}
    \bm{\varLambda}
    \bm{U}^\top
    =
    \sum_{i=1}^N
    \lambda_i
    \bm{u}_i
    \bm{u}_i^\top
    ,
\end{equation}
where
$
\bm{U}
\coloneqq
[
    \bm{u}_1 \; \cdots \; \bm{u}_N
]
\in
\mathbb{R}^{N\times N}
$
is the orthogonal matrix of eigenvectors $\{\bm{u}_i\}_{i=1}^N$,
and
$
\bm{\varLambda}
\coloneqq
\operatorname{diag}
(
    \lambda_1, \ldots, \lambda_N
)
\in
\mathbb{R}_{\geq 0}^{N\times N}
$
is the diagonal matrix of eigenvalues $\{\lambda_i\}_{i=1}^N$.
The eigenvalues are sorted in descending order,
$\lambda_1 \geq \cdots \geq \lambda_N$.
In the end, we perform the projection of the hypercubic vertices
$\bm{s}$ onto
the two-dimensional plane spanned by
the two eigenvectors $\bm{u}_i$ and $\bm{u}_j$.
The projection is given by
\begin{equation}
    \bm{x}_{(i, j)}(\bm{s})
    =
    \begin{bmatrix}
        \bm{u}_i^\top
        \\
        \bm{u}_j^\top
    \end{bmatrix}
    (
        \bm{s} - \langle\bm{s}\rangle
    )
    .
\end{equation}
We use the pair $(i, j) = (1, 2)$.
All four landscapes in Fig.~\ref{fig:fig-01}f share one projection,
which we fit to the example with flippability $\phi = 1$,
so that the four landscapes are directly comparable.
We connect the projected hypercubic vertices if their Hamming distance
\begin{equation}
    D_\mathrm{H}
    (\bm{s}, \bm{s}^\prime)
    \coloneqq
    \sum_{i=1}^{N}
    \bigg(
        1 - \frac{1+s_i s_i^\prime}{2}
    \bigg)
    =
    \frac{N - \bm{s}^\top\bm{s}^\prime}{2}
\end{equation}
is one, i.e., $\bm{s}^\prime = \bm{F}_k \bm{s}$.
We draw hypercubic edges between $\bm{s}$ and $\bm{s}^\prime$ thicker if
$H(\bm{s}) = H(\bm{s}^\prime)$.

For Fig.~\ref{fig:fig-03} in the main text,
the hypercubic vertices are drawn so that their $x$ coordinates are the
Hamming distance from the initial Fock state.
See also ref.~\cite{Horiike2025}.

\section{Robustness to the sign of the couplings}
\label{sec:pmj-supp}
\begin{figure*}[tb]
    \centering
    \includegraphics{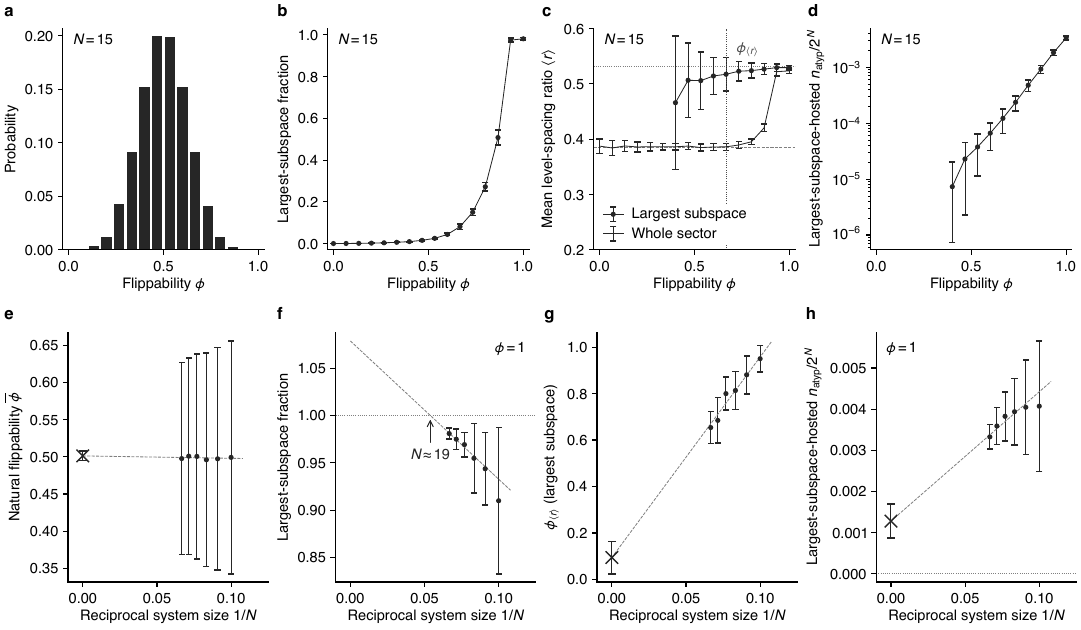}
    \bcaption{
        Flippability, fragmentation and level statistics at $N = 15$,
        and their thermodynamic-limit extrapolation, $\pm J$ ensemble.
    }{
        The $\pm J$ analogue of Fig.~\ref{fig:fig-02} of the main text:
        identical panels and construction, with model $\pm J$ throughout
        \textbf{b}--\textbf{d}.
        Error bars and sample sizes as in Fig.~\ref{fig:fig-02}.
        In \textbf{h},
        the fit does not reach $0$,
        and the cross marks its $N \to \infty$ intercept
        with the standard error of the fit.
    }
    \label{fig:fig-s02}
\end{figure*}

\begin{figure*}[tb]
    \centering
    \includegraphics{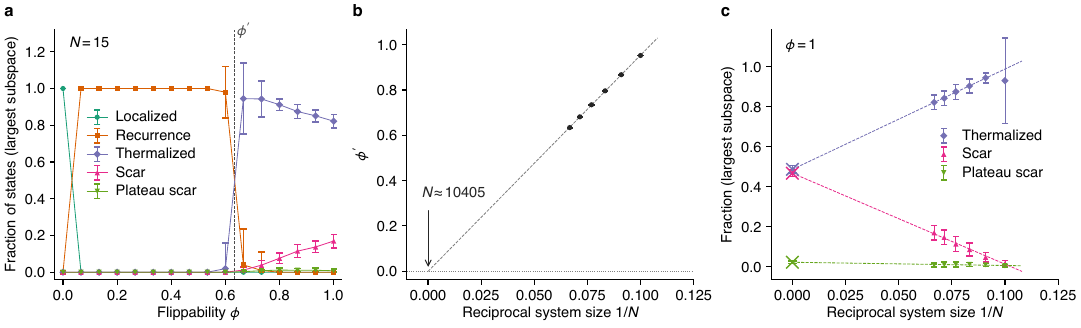}
    \bcaption{
        State counts of the five dynamical classes, and their
        thermodynamic-limit extrapolation, $\pm J$ ensemble.
    }{
        The $\pm J$ analogue of Fig.~\ref{fig:fig-04} of the main text:
        identical panels, classification and extrapolation, with model
        $\pm J$ throughout.
        Error bars as in Fig.~\ref{fig:fig-04}.
    }
    \label{fig:fig-s04}
\end{figure*}

Extended Data Figs.~\ref{fig:fig-s02}--\ref{fig:fig-s04} repeat
Figs.~\ref{fig:fig-02}--\ref{fig:fig-04} of the main text with every
edge coupling drawn independently as $J_{i,j} \in \{+J, -J\}$ (model
$\pm J$) instead of the ferromagnetic $J_{i,j} = +J$, with otherwise
identical construction, classification and extrapolation.
We confirm that the results are qualitatively similar between the
ferromagnetic and $\pm J$ ensembles,
with two quantitative differences in the extrapolations.
The GOE-confident flippability extrapolates to
$\phi_{\langle r \rangle} \to 0.09 \pm 0.07$
(Extended Data Fig.~\ref{fig:fig-s02}g),
compared with $-0.14 \pm 0.09$ in the ferromagnetic ensemble.
The fraction of atypical eigenstates at $\phi = 1$ decreases with $N$
as in the ferromagnetic ensemble,
but its linear extrapolation remains positive,
$n_{\mathrm{atyp}}/2^N \to 0.0013 \pm 0.0004$
(Extended Data Fig.~\ref{fig:fig-s02}h).
Overall,
the agreement indicates that the flippability
characterization is governed by the network/longitudinal-field structure
rather than by the sign of the interactions.

\section{The asymptotic form of the participation entropy}
Here we derive equation~\eqref{eq:s2-eth},
the average participation entropy of a GOE eigenvector of dimension $M$.

\paragraph{Eigenvectors as random unit vectors.}
In the Fock basis, the Hamiltonian is a real symmetric matrix,
so its spectrum follows the GOE and its eigenvectors can be chosen real.
The GOE is invariant under orthogonal transformations,
so an eigenvector $\bm{c} \in \mathbb{R}^M$ is a random unit vector,
uniformly distributed on the sphere.
We generate it as $\bm{c} = \bm{\xi} / |\bm{\xi}|$,
where the components $\xi(\bm{s})$ are independent standard Gaussian
variables;
for large $M$, each amplitude is thus Gaussian with variance $1/M$,
the Porter--Thomas distribution~\cite{Porter1956}.

\paragraph{Moments of the amplitudes.}
The length $|\bm{\xi}|$ and the direction $\bm{c}$ of a Gaussian vector
are independent.
Averaging $\xi(\bm{s}) = |\bm{\xi}|\, c(\bm{s})$ raised to any power
therefore factorizes, for example
$\overline{\xi^4(\bm{s})} = \overline{|\bm{\xi}|^4}\,\overline{c^4(\bm{s})}$,
where $\overline{\,\cdot\,}$ is the ensemble average.
Since $|\bm{\xi}|^2$ is a chi-squared variable with $M$ degrees of
freedom,
$\overline{|\bm{\xi}|^4} = M(M + 2)$ and
$\overline{|\bm{\xi}|^8} = M(M + 2)(M + 4)(M + 6)$.
With the Gaussian moments
$\overline{\xi^4} = 3$ and $\overline{\xi^8} = 105$,
and $\overline{\xi^4(\bm{s})\,\xi^4(\bm{s}^\prime)} = 9$
for $\bm{s} \neq \bm{s}^\prime$,
we obtain
\begin{equation*}
    \overline{c^4(\bm{s})} = \frac{3}{M(M + 2)},
    \quad
    \overline{c^8(\bm{s})} = \frac{105}{M(M + 2)(M + 4)(M + 6)},
    \quad
    \overline{c^4(\bm{s})\,c^4(\bm{s}^\prime)}
    = \frac{9}{M(M + 2)(M + 4)(M + 6)}.
\end{equation*}

\paragraph{Leading order.}
Summing over the $M$ Fock states gives the average inverse
participation ratio,
$\overline{R_2} = M\,\overline{c^4(\bm{s})} = 3/(M + 2)$,
the $q = 2$ case of the exact moments derived in
ref.~\cite{Backer2019}.
$R_2$ is a sum of $M$ weakly correlated terms,
so it barely fluctuates for large $M$,
and $\overline{S_2} = -\overline{\ln (R_2)}$
is close to $-\ln (\overline{R_2})$:
\begin{equation*}
    \overline{S_2} \simeq \ln (M + 2) - \ln (3) \simeq \ln (M) - \ln (3).
\end{equation*}
The offset $\ln (3)$ comes from the Gaussian fourth moment and is
specific to real amplitudes.

\paragraph{Finite-size correction.}
The average of a logarithm is smaller than the logarithm of the average,
so the fluctuations of $R_2$ raise $\overline{S_2}$;
$-\ln(\overline{R_2})$ is therefore only a lower bound on
$\overline{S_2}$~\cite{Backer2019}.
The statistics of eigenfunction amplitudes and of the inverse
participation ratio in random-matrix ensembles are reviewed in
ref.~\cite{Mirlin2000};
the finite-size behaviour of multifractal dimensions of random matrices
is analysed in ref.~\cite{Backer2019},
and the dependence of the participation ratio of many-body eigenstates
on the Hilbert-space dimension in ref.~\cite{Beugeling2015}.
Writing $R_2 = \overline{R_2}\,(1 + \delta)$ with $\overline{\delta} = 0$
and expanding $\ln (1 + \delta) = \delta - \delta^2/2 + \cdots$,
\begin{equation*}
    \overline{S_2}
    =
    -\ln (\overline{R_2}) + \frac{\overline{\delta^2}}{2} + O(M^{-2}),
\end{equation*}
because the higher terms are of order $M^{-2}$.
The relative variance follows from the moments above.
Counting the $M$ diagonal and $M(M - 1)$ off-diagonal pairs in
$R_2^2 = \sum_{\bm{s}, \bm{s}^\prime} c^4(\bm{s})\, c^4(\bm{s}^\prime)$,
\begin{equation*}
    \overline{R_2^2}
    =
    \frac{9M + 96}{(M + 2)(M + 4)(M + 6)},
    \qquad
    \overline{\delta^2}
    =
    \frac{\overline{R_2^2}}{\overline{R_2}^{\,2}} - 1
    =
    \frac{8 (M - 1)}{3 (M + 4)(M + 6)}
    \simeq
    \frac{8}{3M}.
\end{equation*}
Combining the two contributions,
$\ln (M + 2) \simeq \ln (M) + 2/M$ and $\overline{\delta^2}/2 \simeq 4/(3M)$,
gives equation~\eqref{eq:s2-eth},
\begin{equation*}
    \overline{S_2}
    =
    \ln (M) - \ln (3) + \frac{10}{3M} + O(M^{-2}).
\end{equation*}
A direct average over random unit vectors agrees with this expression
to within $0.003$ for $M \geq 100$ and $0.02$ for $M = 30$.

\paragraph{Comparison with the examples.}
Extended Data Fig.~\ref{fig:fig-s05}a--c compares
equation~\eqref{eq:s2-eth} with the eigenstates of the largest kinetically
accessible subspace in the three examples.
We study these specific Hamiltonians,
so we do not add the random longitudinal field $\epsilon_i$ of the main
text.
Instead, we exclude degenerate eigenstates,
whose participation entropy depends on the basis chosen within the
degenerate eigenspace;
for the ring, we exclude only those that remain degenerate after
resolving the momentum and the parity.
The bulk of the eigenstates lies close to the GOE value:
the median of $S_2$ is $5.53$ against $5.81$ in the ferromagnetic
random network ($M = 998$),
$5.17$ against $5.57$ in the $\pm J$ random network ($M = 783$),
and $4.86$ against $4.69$ in the ring ($M = 322$).
The atypical eigenstates in $S_2$ fall far below this value.

\section{Participation entropy and multipartite entanglement}
\label{sec:pe-vs-q-supp}
The main text identifies scar-state candidates by the participation
entropy $S_2$ rather than by entanglement
(\nameref{sec:participation-entropy} in \nameref{sec:methods} in the
main text).
Below we examine and compare the participation entropy with the multipartite
entanglement and explain why we use the participation entropy rather than the
multipartite entanglement.

\subsection{Eigenenergy dependence of multipartite entanglement}
It is well known that the de facto standard for detecting quantum many-body
scar states is checking the entanglement of the
eigenstate~\cite{Moudgalya2022,Chandran2023}.
If the entanglement entropy is anomalously lower than that of typical
eigenstates,
the eigenstate is considered a candidate for a scar state.
It is, however, difficult to apply this entanglement-based criterion
to systems with random interactions for two reasons:
(1) calculating the bipartite entanglement entropy,
which is the de facto standard for measuring entanglement in quantum
many-body systems,
cuts the system into two subsystems,
but it is not obvious how to fairly cut the random interaction network into two
subsystems;
(2) the typical value of entanglement entropy of eigenstates depends
on the corresponding eigenenergy,
making it difficult to systematically identify the atypical
eigenstates in the sector.

We can solve the former problem by using the multipartite (or global)
entanglement entropy~\cite{Meyer2002,Brennen2003,Amico2008},
which is defined by cutting the system into a single spin and the rest
and averaging the linear entanglement entropy over all possible subsystems:
\begin{equation}
    Q(\hat{\rho})
    \coloneqq
    \frac{1}{N}
    \sum_{i=1}^{N}
    Q(\hat{\rho}_i)
    ,
\end{equation}
where
\begin{equation}
    Q(\hat{\rho}_i)
    \coloneqq
    2
    [
        1
        -
        \tr(\hat{\rho}_i^2)
    ]
    \in
    [0, 1]
\end{equation}
is the scaled linear entanglement entropy of spin $i$,
$\hat{\rho} \coloneqq \ket{\psi}\!\bra{\psi}$ is the density matrix,
$\hat{\rho}_i \coloneqq \tr_{\setminus i} (\hat{\rho})$ is the reduced
density matrix of spin $i$,
and $\tr_{\setminus i}$ is the partial trace over all spins except for
spin $i$.
Note that the linear entanglement entropy is the first-order
approximation of the von Neumann entanglement entropy,
$-\tr\bm{(}\hat{\rho}_i \ln (\hat{\rho}_i)\bm{)}$,
in the Mercator expansion.
This measure of entanglement is independent of the partition of the
system and can fairly compare the entanglement across different
random interaction networks.
The multipartite entanglement entropy does not solve the latter problem,
because the eigenenergy bounds $Q(\hat{\rho})$ from above.
Instead, as we have already applied in this work,
we use the participation entropy,
which exhibits behaviour similar to the entanglement
entropy~\cite{Beugeling2015},
but is independent of the eigenenergy value.

Here we derive the upper bound of the multipartite entanglement.
Consider the equienergy sector of classical energy $E_0$,
that is, the Fock states with $H(\bm{s}) = E_0$,
and let $\hat{P}_{E_0}$ be the projector onto it.
A single flip $\bm{s} \to \bm{F}_k \bm{s}$ stays within the sector
only if $H(\bm{F}_k \bm{s}) = H(\bm{s})$,
so the approximated Hamiltonian of
equation~\eqref{eq:approximated-Hamiltonian} restricted to the sector is
\begin{equation}
    \hat{P}_{E_0}
    \hat{\mathcal{H}}
    \hat{P}_{E_0}
    =
    E_0 \hat{P}_{E_0}
    -
    \varGamma
    \sum_{k=1}^N
    \hat{P}_{E_0}
    \hat{X}_k
    \hat{P}_{E_0}
    .
\end{equation}
An eigenstate $\ket{E} = \hat{P}_{E_0} \ket{E}$ with eigenenergy
$E$ therefore satisfies
$E = E_0 - \varGamma \sum_{k=1}^N \braket{\hat{X}_k}$,
where $\braket{\cdot} \coloneqq \braket{E | \cdot | E}$,
so the eigenenergy fixes the spin-averaged transverse magnetization,
\begin{equation}
    \frac{1}{N}
    \sum_{k=1}^N
    \braket{\hat{X}_k}
    =
    -
    \frac{E - E_0}{N \varGamma}
    \label{eq:x-energy}
    .
\end{equation}
The operators
$\hat{\sigma}_0 \coloneqq \hat{1}$,
$\hat{\sigma}_1 \coloneqq \hat{\sigma}_x$,
$\hat{\sigma}_2 \coloneqq \hat{\sigma}_y$,
and
$\hat{\sigma}_3 \coloneqq \hat{\sigma}_z$
form an orthogonal basis of the operator space of a single spin,
$\tr(\hat{\sigma}_u \hat{\sigma}_v) = 2 \delta_{u, v}$,
so that $\hat{\rho}_i = \sum_{w=0}^3 c_w \hat{\sigma}_w / 2$
with some coefficients $c_w$.
Multiplying both sides by $\hat{\sigma}_v$ and taking the trace gives
$\tr(\hat{\rho}_i \hat{\sigma}_v)
= \sum_{w} c_w \tr(\hat{\sigma}_w \hat{\sigma}_v) / 2
= c_v$.
The coefficients are therefore given by
\begin{equation}
    c_w
    =
    \tr(\hat{\rho}_i \hat{\sigma}_w)
    =
    \tr(\hat{\rho} \hat{W}_{i})
    =
    \braket{\hat{W}_{i}}
    ,
\end{equation}
where $\hat{W}_{i} \in \{\hat{X}_i, \hat{Y}_i, \hat{Z}_i\}$ denotes
$\hat{\sigma}_w$ acting on spin $i$.
The second equality follows from the definition of the partial trace:
writing
$\hat{\rho} = \sum_\alpha \hat{A}_\alpha \otimes \hat{B}_\alpha$
with $\hat{A}_\alpha$ acting on spin $i$
and $\hat{B}_\alpha$ acting on the remaining spins,
we have
$\hat{\rho}_i = \sum_\alpha \tr(\hat{B}_\alpha) \hat{A}_\alpha$,
and hence
\begin{align*}
    \tr(\hat{\rho}_i \hat{\sigma}_w)
    & =
    \sum_\alpha
    \tr(\hat{B}_\alpha)
    \tr(\hat{A}_\alpha \hat{\sigma}_w)
    \\
    & =
    \sum_\alpha
    \tr[(\hat{A}_\alpha \hat{\sigma}_w) \otimes \hat{B}_\alpha]
    \\
    & =
    \sum_\alpha
    \tr[
        (\hat{A}_\alpha \otimes \hat{B}_\alpha)
        (\hat{\sigma}_w \otimes \hat{1}_{\setminus i})
    ]
    \\
    & =
    \tr\bigg[
        \bigg(
            \sum_\alpha
            \hat{A}_\alpha \otimes \hat{B}_\alpha
        \bigg)
        (\hat{\sigma}_w \otimes \hat{1}_{\setminus i})
    \bigg]
    \\
    & =
    \tr[\hat{\rho} (\hat{\sigma}_w \otimes \hat{1}_{\setminus i})]
    \\
    & =
    \tr(\hat{\rho} \hat{W}_i)
    .
\end{align*}
The first equality uses the linearity of the trace.
The second uses
$\tr(\hat{P} \otimes \hat{Q}) = \tr(\hat{P}) \tr(\hat{Q})$
read from right to left.
The third uses the mixed-product property
$(\hat{P} \hat{R}) \otimes (\hat{Q} \hat{S})
= (\hat{P} \otimes \hat{Q}) (\hat{R} \otimes \hat{S})$
with $\hat{R} = \hat{\sigma}_w$ and $\hat{S} = \hat{1}_{\setminus i}$.
The fourth uses linearity again,
and the last line recognizes the decomposition of $\hat{\rho}$
and the definition of $\hat{W}_i$.
Since $c_0 = \tr(\hat{\rho}_i) = 1$,
the reduced density matrix of a single spin is
\begin{equation}
    \hat{\rho}_i
    =
    \frac{1}{2}
    (
        \hat{1}
        +
        \braket{\hat{X}_i} \hat{X}
        +
        \braket{\hat{Y}_i} \hat{Y}
        +
        \braket{\hat{Z}_i} \hat{Z}
    )
    ,
\end{equation}
which gives
\begin{equation}
    \tr(\hat{\rho}_i^2)
    =
    \frac{1}{2}
    (
        1
        +
        \braket{\hat{X}_i}^2
        +
        \braket{\hat{Y}_i}^2
        +
        \braket{\hat{Z}_i}^2
    )
    ,
\end{equation}
and
\begin{equation}
    Q(\hat{\rho}_i)
    =
    1
    -
    \braket{\hat{X}_i}^2
    -
    \braket{\hat{Y}_i}^2
    -
    \braket{\hat{Z}_i}^2
    \leq
    1
    -
    \braket{\hat{X}_i}^2
    \label{eq:q-bloch}
    .
\end{equation}
The equality~\cite{Amico2008} is the one-tangle of spin $i$,
so $Q(\hat{\rho})$ is the one-tangle averaged over the spins.
Averaging over spins and using the Cauchy--Schwarz inequality,
$
(1/N) \sum_{i} \braket{\hat{X}_i}^2
\geq
[(1/N) \sum_{i} \braket{\hat{X}_i}]^2
$,
together with equation~\eqref{eq:x-energy},
we obtain
\begin{equation}
    Q(\hat{\rho})
    \leq
    1
    -
    \bigg(
        \frac{E - E_0}{N \varGamma}
    \bigg)^2
    \label{eq:q-energy-bound}
    .
\end{equation}
The bound is saturated when $\braket{\hat{Y}_i} = \braket{\hat{Z}_i} = 0$
and $\braket{\hat{X}_i}$ is uniform over the spins.
The multipartite entanglement is thus suppressed toward both edges of
the spectrum regardless of how delocalized the eigenstate is
(Extended Data Fig.~\ref{fig:fig-s05}d--f),
whereas $S_2$ has no such energy envelope
(Extended Data Fig.~\ref{fig:fig-s05}a--c).

In the two random networks,
$S_2$ and $Q$ are positively correlated
($r = 0.56$ and $0.65$),
and the correlation strengthens once the energy envelope is regressed
out
($r_{|E} = 0.73$ and $0.72$,
Extended Data Fig.~\ref{fig:fig-s05}j,k).
The atypical eigenstates in $S_2$ are therefore also weakly entangled.
In the ring,
the raw correlation is negative,
because $Q$ is dominated by the envelope of
equation~\eqref{eq:q-energy-bound}:
the atypical eigenstates in $S_2$ sit near the center of the spectrum,
where the bound is loosest,
and their $Q$ scatters both above and below that of typical
eigenstates at the same energy (Extended Data Fig.~\ref{fig:fig-s05}f,l).
Without the envelope the correlation is weak.
The multipartite entanglement alone does not single out the PXP scar
states,
whereas the participation entropy does (Extended Data Fig.~\ref{fig:fig-s05}c).

\subsection{Half-system entanglement entropy and multipartite entanglement}
The quantum many-body scars of the PXP model were identified by their
anomalously low bipartite entanglement entropy~\cite{Turner2018,Turner2018a}.
The multipartite entanglement $Q$,
however,
does not single out the atypical eigenstates in $S_2$ of the
ring
(Extended Data Fig.~\ref{fig:fig-s05}f,l).
We therefore compare $Q$ with the von Neumann entanglement entropy of
half of the system,
\begin{equation}
    S_{\mathrm{half}} \coloneqq -\tr\bm{(}\hat{\rho}_A \ln (\hat{\rho}_A)\bm{)}
    ,
\end{equation}
where $\hat{\rho}_A$ is the reduced density matrix of a set $A$ of
$N/2$ spins.
For the ring,
$A$ is a contiguous half of the ring.
All such cuts give the same $S_{\mathrm{half}}$ for an eigenstate of
the translation by one site,
as are the eigenstates retained above;
we average $S_{\mathrm{half}}$ over the $N/2 = 6$ distinct
contiguous cuts.
A random network has no spatial structure that singles out a cut,
so we evaluate all $\binom{N}{N/2}/2 = 462$ balanced bipartitions
and use the one whose $S_{\mathrm{half}}$ averaged over the sector is
the median over the bipartitions.
The sector average depends only weakly on the bipartition
(Extended Data Fig.~\ref{fig:fig-s06}a,b).
The correlation between $S_{\mathrm{half}}$ and $Q$ is likewise
insensitive to the bipartition
(Extended Data Fig.~\ref{fig:fig-s06}c--f).

We first check that the ring reproduces the anomalously
low entanglement eigenstates reported by
Turner \textit{et al.}~\cite{Turner2018a}.
The sector of the N\'eel state of the ring has the dimension
$M = 322$ of the PXP model on $L = 12$ sites with periodic boundary
conditions,
and its Hamiltonian is $E_0 - \varGamma$ times the PXP Hamiltonian.
Resolving the crystal momentum $\kappa$
(the eigenstates of the PXP model on a ring are also eigenstates of the
    translation by one site, with eigenvalue $\exp(\mathrm{i}\kappa)$ and
$\kappa = 2\pi m/L$ for $m = 0, 1, \dots, L - 1$),
we find $L + 1 = 13$ special eigenstates with low $S_{\mathrm{half}}$
and high overlap with the N\'eel state
(Extended Data Fig.~\ref{fig:fig-s06}g,h).
The results agree with those of ref.~\cite{Turner2018a}.

Extended Data Fig.~\ref{fig:fig-s05}g--i plots $S_{\mathrm{half}}$ against the
eigenenergy,
and Extended Data Fig.~\ref{fig:fig-s05}m--o plots it against $Q$,
for the same eigenstates as in Extended Data Fig.~\ref{fig:fig-s05}a--f.
To identify atypical eigenstates in $S_{\mathrm{half}}$ at a given
energy,
we compare each eigenstate with the other eigenstates within
$|E^\prime - E| < 0.5\varGamma$
and flag it when its $S_{\mathrm{half}}$ is more than $3$ standard
deviations below their mean,
as Turner \textit{et al.}~\cite{Turner2018a} identified the scars as
atypical eigenstates relative to eigenstates at the same energy.
To separate the correlation between two quantities $A$ and $B$ of the
eigenstates from their common dependence on the eigenenergy,
we use, in addition to the Pearson correlation coefficient $r(A, B)$
over the eigenstates of the sector,
the partial correlation coefficient
\begin{equation}
    r_{|E}(A, B)
    \coloneqq
    r\big(A - \tilde{A}, B - \tilde{B}\big)
    \label{eq:partial-correlation}
    ,
\end{equation}
where $\tilde{A} = a_0 + a_1 x^2$ is the least-squares fit of $A$ over
the eigenstates of the sector with $x = (E - E_0)/(N\varGamma)$,
the variable of equation~\eqref{eq:q-energy-bound},
and likewise for $\tilde{B}$.
In the two random networks,
$S_{\mathrm{half}}$ and $Q$ are almost linearly related
($r = 0.99$ and $0.98$ and $r_{|E} = 0.95$ and $0.90$,
Extended Data Fig.~\ref{fig:fig-s05}m,n),
and the eigenstates with anomalously low $S_{\mathrm{half}}$ also have
anomalously low $Q$.
The two measures of entanglement are effectively equivalent,
and both correlate with the participation entropy
($r = 0.57$ and $0.71$ between $S_{\mathrm{half}}$ and $S_2$).
Most eigenstates with anomalously low $S_{\mathrm{half}}$
are also atypical in $S_2$:
all $8$ in the ferromagnetic network and $10$ of $14$ in the $\pm J$
network.

In the ring,
the correlation between $S_{\mathrm{half}}$ and $Q$
($r = 0.72$)
arises from the dependence of both on the eigenenergy:
once the quadratic dependence on $E - E_0$ is regressed out of each,
the partial correlation of equation~\eqref{eq:partial-correlation} is
$r_{|E} = -0.18$
(Extended Data Fig.~\ref{fig:fig-s05}o).
The six states of the scar tower outside the zero-energy manifold,
at $(E - E_0)/\varGamma = \pm 1.34$, $\pm 2.67$, and $\pm 3.95$,
have $S_{\mathrm{half}}$ between $1.7$ and $2.7$ standard deviations
below that of eigenstates at the same eigenenergy,
although none of them crosses the cut of $3$ standard deviations.
The special states nearer the edges of the spectrum have fewer than ten
eigenstates within the energy window $|E^\prime - E| < 0.5\varGamma$,
and we do not assess them.
The multipartite entanglement $Q$ of the six special states,
in contrast,
is between $2.2$ and $2.6$ standard deviations \emph{above} that of
eigenstates at the same eigenenergy.
The multipartite entanglement $Q$ thus classifies the PXP scars as more
entangled than typical eigenstates,
whereas the half-system entanglement entropy classifies them as less
entangled.

This difference has a simple origin.
According to the eigenstate thermalization
hypothesis~\cite{Deutsch1991,Srednicki1994,Rigol2008,DAlessio2016},
the reduced density matrix of a small subsystem in an eigenstate
$\ket{E}$ coincides with that of the canonical ensemble at the
temperature set by $E$.
The expectation values $\braket{\hat{X}_i}$, $\braket{\hat{Y}_i}$,
and $\braket{\hat{Z}_i}$ in equation~\eqref{eq:q-bloch},
and hence $Q(\hat{\rho})$,
therefore take their thermal values,
and $S_{\mathrm{half}}$ takes the thermal entropy of half of the
system;
both are largest at the center of the spectrum of the sector.
The two quantities are also related by an inequality~\cite{Amico2008}.
The eigenvalues of $\hat{\rho}_i$ are
$\lambda_\pm = [1 \pm \sqrt{1 - Q(\hat{\rho}_i)}]/2$
because
$\tr(\hat{\rho}_i)=1$ and $\tr(\hat{\rho}_i^2)=1-Q(\hat{\rho}_i)/2$
give
$\lambda_+ + \lambda_- = 1$ and
$\lambda_+\lambda_- = [1-\tr(\hat{\rho}_i^2)]/2 = Q(\hat{\rho}_i)/4$,
so the von Neumann entanglement entropy of spin $i$,
$-\tr\bm{(}\hat{\rho}_i \ln (\hat{\rho}_i)\bm{)}
= \eta\bm{(}Q(\hat{\rho}_i)\bm{)}$
with $\eta(Q) \coloneqq -\sum_{\pm} \lambda_\pm \ln (\lambda_\pm)$,
is an increasing concave function of $Q(\hat{\rho}_i)$.
For a pure state,
a set $A$ of $N/2$ spins and its complement $\bar{A}$ have the same
von Neumann entanglement entropy $S_{\mathrm{half}}$.
Applying the subadditivity of the von Neumann entropy to both $A$ and
$\bar{A}$,
the concavity of $\eta$,
and equation~\eqref{eq:q-energy-bound} in turn,
we obtain
\begin{equation}
    S_{\mathrm{half}}
    \leq
    \frac{1}{2}
    \sum_{i=1}^{N}
    \eta\bm{(}Q(\hat{\rho}_i)\bm{)}
    \leq
    \frac{N}{2}
    \eta\bm{(}Q(\hat{\rho})\bm{)}
    \leq
    \frac{N}{2}
    \eta
    \Bigg(
        1
        -
        \bigg(
            \frac{E - E_0}{N \varGamma}
        \bigg)^2
    \Bigg)
    \label{eq:shalf-q-bound}
    .
\end{equation}
A low $Q(\hat{\rho})$ thus forces a low $S_{\mathrm{half}}$,
but a high $Q(\hat{\rho})$ does not force a high $S_{\mathrm{half}}$.
The PXP scars approximately form the multiplet of an emergent large
collective spin~\cite{Choi2019}.
Such collective states are nearly maximally mixed on each single spin
yet carry little entanglement between the two halves of the system.
This can be understood through the following example:
the Dicke state of $N = 12$ spins with $N/2$ spins up,
\begin{align*}
    \ket{\mathrm{Dicke}_{N}^{N/2}}
    =
    \binom{N}{N/2}^{-1/2}
    \sum_{\bm{s} \in \{+1, -1\}^N : \sum_{i=1}^{N} s_i = 0}
    \ket{\bm{s}}
    ,
\end{align*}
has
$Q(\hat{\rho}) = 1$ but $S_{\mathrm{half}} = 1.32$,
far below the maximal value $(N/2) \ln (2) = 4.16$.
Because $Q(\hat{\rho})$ probes only the reduced density matrices of
single spins,
it cannot distinguish such collective states from thermal ones.
The half-system entanglement entropy $S_{\mathrm{half}}$ and the
participation entropy $S_2$ both detect the PXP scars,
but through different mechanisms:
$S_{\mathrm{half}}$ through the entanglement between the two halves of
the system,
and $S_2$ through the localization in the Fock-space landscape.
Indeed,
$S_{\mathrm{half}}$ and $S_2$ are uncorrelated over the sector of the
ring ($r = -0.01$).
Of the five atypical eigenstates in $S_2$,
four are the special states at $(E - E_0)/\varGamma = \pm 1.34$
and $\pm 2.67$,
which are also low in $S_{\mathrm{half}}$,
whereas the fifth lies on the zero-energy manifold and has a typical
$S_{\mathrm{half}}$.

\subsection{Time evolution of participation entropy and entanglement}
Extended Data Fig.~\ref{fig:fig-s07} shows the time evolution of $S_2$, $Q$,
and $S_{\mathrm{half}}$
for the same initial Fock states as Fig.~\ref{fig:fig-03} of the main
text and Extended Data Fig.~\ref{fig:fig-s03}.
Because $\braket{\bm{s}|\hat{X}_k|\bm{s}} = 0$ for every Fock state
$\ket{\bm{s}}$,
the energy expectation of the initial state is $E_0$,
and energy conservation keeps
$\sum_{k} \braket{\hat{X}_k(t)} = 0$
[equation~\eqref{eq:x-energy} with $E$ replaced by
$\braket{\hat{H}} = E_0$].
The energy bound of equation~\eqref{eq:q-energy-bound} therefore
reduces to the trivial $Q(t) \leq 1$;
$Q(t)$ is limited only by the individual magnetizations in
equation~\eqref{eq:q-bloch}.
The localized state stays at $S_2 = Q = 0$,
and the recurrence state returns to $S_2 = Q = 0$ periodically.
The thermalized states reach their saturation values within
$Jt \approx 50$.
In the random networks,
$Q(t)$ of a thermalized state saturates close to $1$.
In the ring,
it saturates at $Q \approx 0.78$,
because the sector carries a net magnetization,
$(1/N) \sum_{i} \braket{\hat{Z}_i} \approx -0.45$,
and $Q \leq 1 - (1/N) \sum_{i} \braket{\hat{Z}_i}^2$.
The scar and plateau-scar states of the random networks spread over
the Fock-space landscape slowly,
and both $S_2(t)$ and $Q(t)$ remain below their thermalized values
over the whole time window.
The N\'eel scar of the ring behaves differently.
Its $S_2(t)$ oscillates between nearly $0$ and the thermalized value,
dropping each time the wave packet refocuses on a few Fock states,
whereas $Q(t)$ grows smoothly with only small oscillations.
This is the dynamical counterpart of
Extended Data Fig.~\ref{fig:fig-s05}c,f,l:
the scar dynamics is visible in the participation entropy but hardly
in the single-spin quantity $Q$.

Extended Data Fig.~\ref{fig:fig-s07}c,f shows $S_{\mathrm{half}}(t)$
with the same
bipartitions as Extended Data Fig.~\ref{fig:fig-s05}:
the median balanced bipartition for the random networks
and the average over the $N/2 = 6$ contiguous half-ring cuts for the
rings.
The thermalized states saturate within $Jt \approx 50$.
The scar and plateau-scar states of the random networks entangle the
two halves more slowly,
and the plateau-scar states saturate below the thermalized value.
For both classes,
$S_{\mathrm{half}}(t)$ follows $Q(t)$ closely.
Unlike $S_2$ and $Q$,
$S_{\mathrm{half}}$ depends on the bipartition.
In the recurrence states,
two spins flip back and forth while the others stay frozen.
In the ferromagnetic network the two spins lie on the same side of the
cut,
so $S_{\mathrm{half}}(t) = 0$ while $S_2(t)$ and $Q(t)$ oscillate;
in the $\pm J$ network they lie on opposite sides,
and $S_{\mathrm{half}}(t)$ oscillates between $0$ and $0.25$.
The N\'eel scar of the ring grows its half-system entanglement slowly,
with small oscillations,
reaching only $S_{\mathrm{half}} \approx 1.2$,
half of the thermalized value of the ring,
in agreement with the slowest entanglement growth from the N\'eel
state of the PXP model~\cite{Turner2018}.
The late-time value of the N\'eel scar relative to the thermalized
state is $0.51$ for $S_{\mathrm{half}}$
but $0.73$ for $Q$ ($0.56$ against $0.78$):
as for the eigenstates (Extended Data Fig.~\ref{fig:fig-s05}i,o),
the half-system entanglement entropy separates the N\'eel scar from
thermalization more clearly than $Q$.
The $\pm J$ ring gives curves identical to those of the ferromagnetic
ring,
because its initial states are the gauge images of those of the
ferromagnetic ring,
and the gauge transformation permutes Fock states and acts on each spin
locally,
leaving $S_2$, $Q$, and $S_{\mathrm{half}}$ unchanged.

\section{Calibration of the five-class thresholds}
\label{sec:threshold-calibration}
The five-class classification of
\nameref{sec:classification} in the main text
classifies every initial Fock state $\ket{\bm{s}_0}$ into a class from its
fidelity $F(t; \bm{s}_0)$,
applying four tests in the order of the flowchart in
Fig.~\ref{fig:fig-03}a of the main text.
A state is localized if $1 - \min_t F(t; \bm{s}_0) < 0.1$.
Otherwise,
it is a recurrence if its subspace is smaller than the recurrence
threshold;
it is thermalized if its anomaly
$F_\mathrm{rev}(\bm{s}_0) / \med_{\bm{s}_0}\bm{(}F_\mathrm{rev}(\bm{s}_0)\bm{)}$
is below $3$;
and it is a scar or a plateau scar if its floor ratio
$\min_t F(t; \bm{s}_0) / F_\mathrm{rev}(\bm{s}_0)$
is below or above $0.1$.
Two of these thresholds,
the recurrence threshold and the anomaly threshold,
are calibrated against a random-matrix null model
(Extended Data Fig.~\ref{fig:fig-s08}a--d).
The other two,
the localization and floor thresholds,
are conventions on the same $0.1$ scale,
and we show that the data offer no natural alternative
(Extended Data Fig.~\ref{fig:fig-s08}e).

\subsection{Definition of revival}
We measure the revival after the thermalized states have relaxed.
We define the revival $F_\mathrm{rev}(\bm{s}_0)$ as the largest
fidelity after $t_\mathrm{relax} = 50/J$,
by which $S_2(t)$, $Q(t)$, and $S_{\mathrm{half}}(t)$ of the
thermalized states saturate (Extended Data Fig.~\ref{fig:fig-s07}).
The alternative,
the largest fidelity after the deepest dip $t_{\min}$ in an early
interval,
fails for either length of that interval.
A long interval,
such as $\tau/4$ with $\tau = 2000/J$,
places $t_{\min}$ on a near-zero dip after the first revival,
which then no longer counts.
An interval ending at $t_\mathrm{relax}$ counts a side lobe of the
initial decay as a revival:
the $99$th percentile of the GOE null below then levels off near the
first side lobe of the semicircle decay,
$0.0175$,
while its median falls to $0.004$.
On the $N = 12$ ferromagnetic and $\pm J$ models
($32$ networks each at $\phi = 0.25$, $0.5$, $0.75$, and $1$),
our definition leaves the localized and recurrence states unchanged
and raises the scars from $102$ to $161$ and from $892$ to $1251$
relative to the quarter-window rule.
The added scars relax fully before $Jt = 50$ and peak at a median
$Jt \approx 60$.

\subsection{Typical fidelity revival of GOE matrices}
We tabulate the revival that a Fock-initialized state reaches by chance
under a GOE Hamiltonian,
because a revival has no absolute scale:
a state returns closer to its initial state in a smaller subspace.
For $38$ logarithmically spaced dimensions $2 \leq D \leq 4096$,
we draw $(\hat{G} + \hat{G}^\top)/2$,
with $\hat{G}$ a $D \times D$ matrix of independent standard normal
entries,
and evolve a random basis state.
We rescale the spectrum so that its bandwidth
$W \coloneqq \max_n E_n - \min_n E_n$
equals the median bandwidth of kinetically accessible subspaces of the
same dimension in real networks.
This median comes from the largest equienergy sector of $12$
ferromagnetic networks with $N = 13$ at every flippability level,
covering $216$ dimensions up to $2037$;
we interpolate it logarithmically and hold it fixed above $2037$.
The time step $0.8\pi/W$,
the window $\tau = 2000/J$,
and $t_\mathrm{relax} = 50/J$ follow the main text.
We draw $200$ samples for $D \leq 64$ and $\max(20, 12800/D)$ above,
and record the median and the $90$th and $99$th percentiles of
$F_\mathrm{rev}(\bm{s}_0)$.
The bandwidth matters:
a wider spectrum completes more oscillations within the fixed window
and so has more chances to revive.
The grid density hardly matters:
at $D = 32$,
the null median is $0.415$ to $0.446$ for $400$ to $3200$ grid points,
and the grid used has $605$.

\subsection{Recurrence threshold}
We call a state a recurrence if its subspace is so small that a random
state revives as strongly.
The recurrence threshold is the dimension $D^\ast$ at which the null
median of $F_\mathrm{rev}(\bm{s}_0)$ crosses $0.1$,
the scale of the localization threshold.
The null median falls from $0.986$ at $D = 4$ to $0.237$ at $D = 68$
and $0.047$ at $D = 392$ (Extended Data Fig.~\ref{fig:fig-s08}a),
crossing $0.1$ at $D^\ast = 168$,
which we round to $D \lesssim 180$.
The mean level-spacing ratios of these small subspaces lie between the
Poisson and GOE values,
$\langle r \rangle = 0.449$ at $D = 8$--$12$ and $0.480$ at
$D = 32$--$64$.
We compare with the null rather than with the median revival of the
state's own subspace,
which misfires on constrained spectra:
the $322$-state sector of the one-dimensional ring has a high median
revival and would make even the N\'eel scar a recurrence.

\subsection{Anomaly threshold}
\begin{figure*}[tb]
    \centering
    \includegraphics{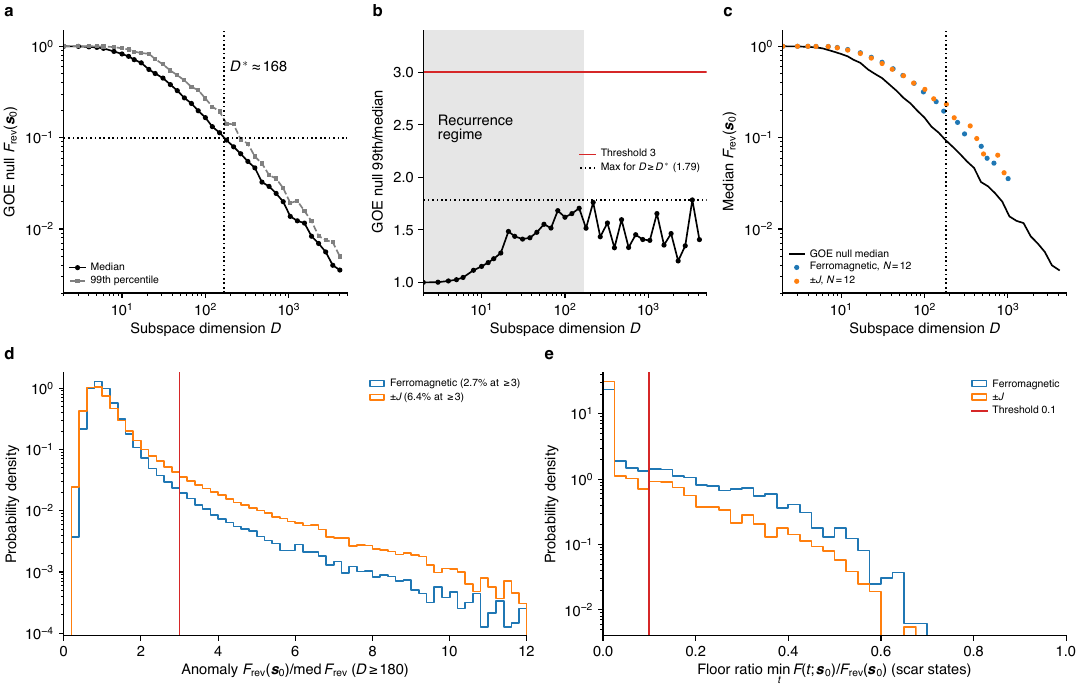}
    \bcaption{
        Calibration of the five-class thresholds.
    }{
        \textbf{a}--\textbf{c},
        GOE revival null:
        a random basis state evolved under a GOE matrix whose spectrum is
        rescaled to the median bandwidth of the real subspaces of the
        same dimension $D$,
        with the time grid and the revival
        $F_\mathrm{rev}(\bm{s}_0)$ of the main text.
        \textbf{a},
        Median (black) and $99$th percentile (grey) of the null
        $F_\mathrm{rev}(\bm{s}_0)$ against $D$.
        Dotted lines,
        $F_\mathrm{rev}(\bm{s}_0) = 0.1$ and its crossing with the median,
        $D^\ast \approx 168$,
        which sets the recurrence threshold $D \lesssim 180$.
        \textbf{b},
        Ratio of the $99$th percentile of the null to its median.
        Shaded,
        the recurrence regime $D < D^\ast$.
        Red line,
        the anomaly threshold $3$;
        dotted line,
        the largest ratio for $D \geq D^\ast$, $1.79$.
        \textbf{c},
        Median $F_\mathrm{rev}(\bm{s}_0)$ of all initial Fock states of
        the $N = 12$ ferromagnetic (blue) and $\pm J$ (orange) ensembles
        of Fig.~\ref{fig:fig-04} of the main text,
        binned by $D$,
        compared with the null median (black line).
        Dotted line,
        $D = 180$.
        \textbf{d},
        Distribution of the anomaly
        $F_\mathrm{rev}(\bm{s}_0) /
        \med_{\bm{s}_0}\bm{(}F_\mathrm{rev}(\bm{s}_0)\bm{)}$
        of the states with $D \geq 180$ in the same ensembles.
        Red line,
        the anomaly threshold $3$;
        the legend gives the fraction of states at or above it.
        \textbf{e},
        Distribution of the floor ratio
        $\min_t F(t; \bm{s}_0) / F_\mathrm{rev}(\bm{s}_0)$
        of the states classified as scars or plateau scars in the same
        ensembles.
        Red line,
        the floor threshold $0.1$.
    }
    \label{fig:fig-s08}
\end{figure*}

Above the recurrence threshold,
we require a scar to revive at least three times more than the typical
state of its own subspace.
The anomaly divides $F_\mathrm{rev}(\bm{s}_0)$ by the median over
states of the same dimension in the same realization,
pooling equal-sized subspaces,
which are statistically interchangeable.
Chance alone places the $99$th percentile of the null only $1.20$ to
$1.79$ times above its median for $D \geq D^\ast$
(Extended Data Fig.~\ref{fig:fig-s08}b),
so the threshold $3$ leaves a margin of $1.7$.
The null itself cannot serve as the reference:
in the $N = 12$ ensembles of Fig.~\ref{fig:fig-04} of the main text,
the real median exceeds the null median by $1.29$ at $D \approx 16$
and $2.42$ at $D \approx 1000$ in the ferromagnetic model,
and by $1.27$ and $2.28$ in the $\pm J$ model
(Extended Data Fig.~\ref{fig:fig-s08}c),
because sparse graphs recur more than random matrices.
The anomaly of states with $D \geq 180$ peaks near $1$ with a long
tail (Extended Data Fig.~\ref{fig:fig-s08}d),
and the threshold selects $2.7\%$ of them in the ferromagnetic model
and $6.4\%$ in the $\pm J$ model,
including the oscillating and plateau scars of Fig.~\ref{fig:fig-03}
of the main text.

\subsection{Localization and floor thresholds}
We set the localization and floor thresholds by convention on the same
$0.1$ scale.
A state is localized if it never leaves its initial Fock state by more
than $10\%$,
$1 - \min_t F(t; \bm{s}_0) < 0.1$.
The floor ratio $\min_t F(t; \bm{s}_0) / F_\mathrm{rev}(\bm{s}_0)$
splits the anomalous states into scars,
which relax fully between revivals,
and plateau scars,
which stay elevated after the first collapse.
For the $6461$ ferromagnetic and $14610$ $\pm J$ anomalous states,
the floor ratio peaks at $0$ and decreases toward its largest values,
$0.69$ and $0.66$,
without a gap that would single out a threshold
(Extended Data Fig.~\ref{fig:fig-s08}e).
The threshold $0.1$ classifies $71\%$ and $86\%$ of them as scars.

\section{Supplementary Movies}

\noindent\textbf{Supplementary Movie 1 \textbar{} Time evolution of the
dynamical classes of Fig.~\ref{fig:fig-03}.}
Fig.~\ref{fig:fig-03}b--v of the main text,
animated over $0 \leq Jt \leq 400$.
For the five initial Fock states of the random network
(localized, recurrence, thermalized, scar, plateau scar)
and the two ring states (staggered scar, thermalizing),
the fidelity $F(t)$,
the magnetization per spin $m(t)$,
and the spin expectation values $\langle \hat{Z}_i(t) \rangle$
are drawn up to the current time.
The Fock-space landscapes are coloured by the instantaneous occupation
probability on a logarithmic scale.
Layout and colour conventions are as in Fig.~\ref{fig:fig-03}.
Duration $60$~s at $30$ frames per second.

\begin{figure*}[tb]
    \centering
    \includegraphics{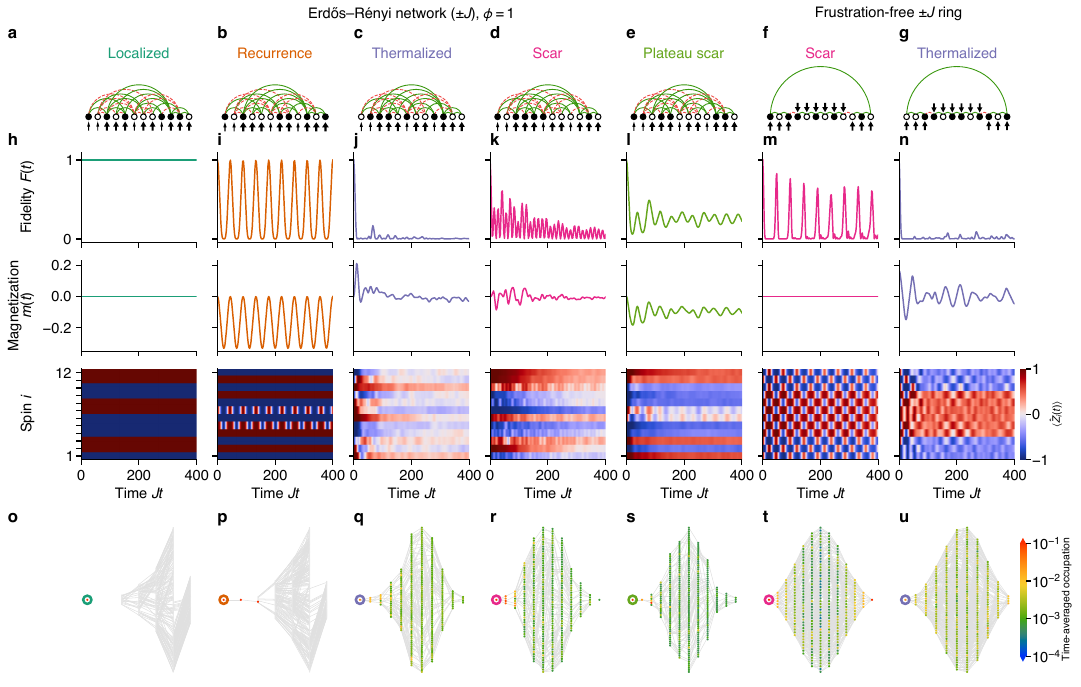}
    \bcaption{
        Dynamical classes on one $\pm J$ network and one sector, and on
        a frustration-free $\pm J$ ring.
    }{
        The $\pm J$ analogue of Fig.~\ref{fig:fig-03} of the main text.
        \textbf{b}--\textbf{v},
        One $\pm J$ realization, $N = 12$ at $\phi = 1$, with five
        initial Fock states of zero total magnetization
        (localized, recurrence, thermalized, scar, plateau scar)
        and a Mattis-gauge $\pm J$ ring
        (staggered scar and a thermalizing Fock state,
        the gauge images of the ring states of Fig.~\ref{fig:fig-03}).
        Network selection as in Methods of the main text
        (spectral-purity ranking among qualifying $\pm J$ networks).
        Panel layout, colour and marker conventions as in Fig.~\ref{fig:fig-03}:
        \textbf{i}--\textbf{o} show the fidelity $F(t)$,
        the magnetization per spin $m(t)$
        and the spin expectation value $\langle \hat{Z}_i(t) \rangle$;
        \textbf{p}--\textbf{v} show the $322$ most-occupied Fock states,
        the dimension of the ring sector.
        Field arrows on the ring follow the sign of the
        Mattis-gauge-transformed longitudinal field.
    }
    \label{fig:fig-s03}
\end{figure*}

\begin{figure*}[tb]
    \centering
    \includegraphics{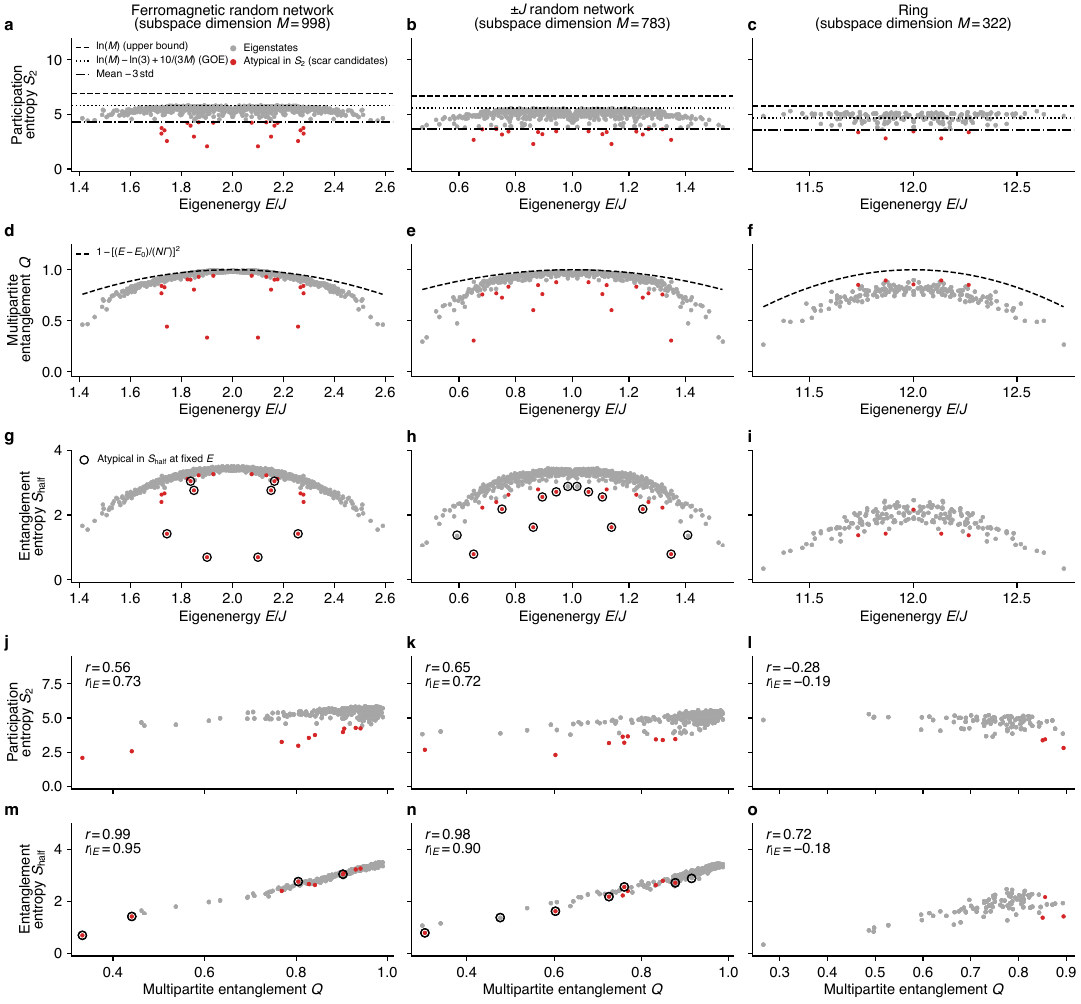}
    \bcaption{
        Participation entropy and multipartite entanglement of the
        eigenstates.
    }{
        Nondegenerate eigenstates of the largest kinetically accessible
        subspace of the sector Hamiltonian
        ($\varGamma = 0.1J$, $N = 12$) for the three examples of
        Fig.~\ref{fig:fig-03} of the main text and
        Extended Data Fig.~\ref{fig:fig-s03}.
        \textbf{a}, \textbf{d}, \textbf{g}, \textbf{j}, \textbf{m},
        Ferromagnetic random network
        (subspace dimension $M = 998$).
        \textbf{b}, \textbf{e}, \textbf{h}, \textbf{k}, \textbf{n},
        $\pm J$ random network
        (subspace dimension $M = 783$).
        \textbf{c}, \textbf{f}, \textbf{i}, \textbf{l}, \textbf{o},
        Ring with uniform field $h = 2J$,
        sector of the N\'eel Fock state
        (dimension $M = 322$).
        \textbf{a}--\textbf{c},
        Rényi-2 participation entropy $S_2$ versus $E$.
        Dashed line,
        the strict upper bound $S_2 \leq \ln(M)$,
        reached only by the uniform superposition over the subspace of
        dimension $M$.
        Dotted line,
        the GOE value $\ln(M) - \ln(3) + 10/(3M)$ of
        equation~\eqref{eq:s2-eth} of the main text.
        Dash-dotted line,
        the cut for atypical eigenstates in $S_2$,
        $3$ standard deviations below the average of $S_2$ over the
        plotted eigenstates.
        \textbf{d}--\textbf{f},
        Multipartite entanglement $Q$ versus eigenenergy $E$.
        Dashed curve,
        the bound of equation~\eqref{eq:q-energy-bound}.
        \textbf{g}--\textbf{i},
        Half-system von Neumann entanglement entropy $S_{\mathrm{half}}$
        versus $E$.
        \textbf{j}--\textbf{l},
        $S_2$ versus $Q$.
        $r$ is the Pearson correlation coefficient,
        and $r_{|E}$ is the partial correlation coefficient of
        equation~\eqref{eq:partial-correlation},
        after regressing both quantities on
        $[(E - E_0)/(N \varGamma)]^2$.
        \textbf{m}--\textbf{o},
        $S_{\mathrm{half}}$ versus $Q$,
        with $r$ and $r_{|E}$ as in \textbf{j}--\textbf{l}.
        Panels in the same column share the $Q$ axis of \textbf{j}--\textbf{o}.
        For the ring, $S_{\mathrm{half}}$ is averaged over the $6$
        contiguous half-ring cuts;
        for the random networks, it is evaluated for the balanced
        bipartition whose sector-averaged $S_{\mathrm{half}}$ is the
        median over all $462$ balanced bipartitions.
        Open circles,
        atypical eigenstates in $S_{\mathrm{half}}$,
        whose $S_{\mathrm{half}}$ is more than $3$ standard
        deviations below the mean of the eigenstates within
        $\pm 0.5\varGamma$ in energy.
        Red points,
        atypical eigenstates in $S_2$ (scar-state candidates):
        $20$ in the ferromagnetic network,
        $18$ in the $\pm J$ network,
        and $5$ in the ring.
        No random longitudinal field is added.
        Degenerate eigenstates are excluded,
        leaving $958$, $746$, and $308$ of the $M$ eigenstates;
        for the ring,
        the translation by one site and the reflection are resolved
        first,
        and only the eigenstates degenerate within the same momentum
        and parity are excluded.
    }
    \label{fig:fig-s05}
\end{figure*}

\begin{figure*}[tb]
    \centering
    \includegraphics{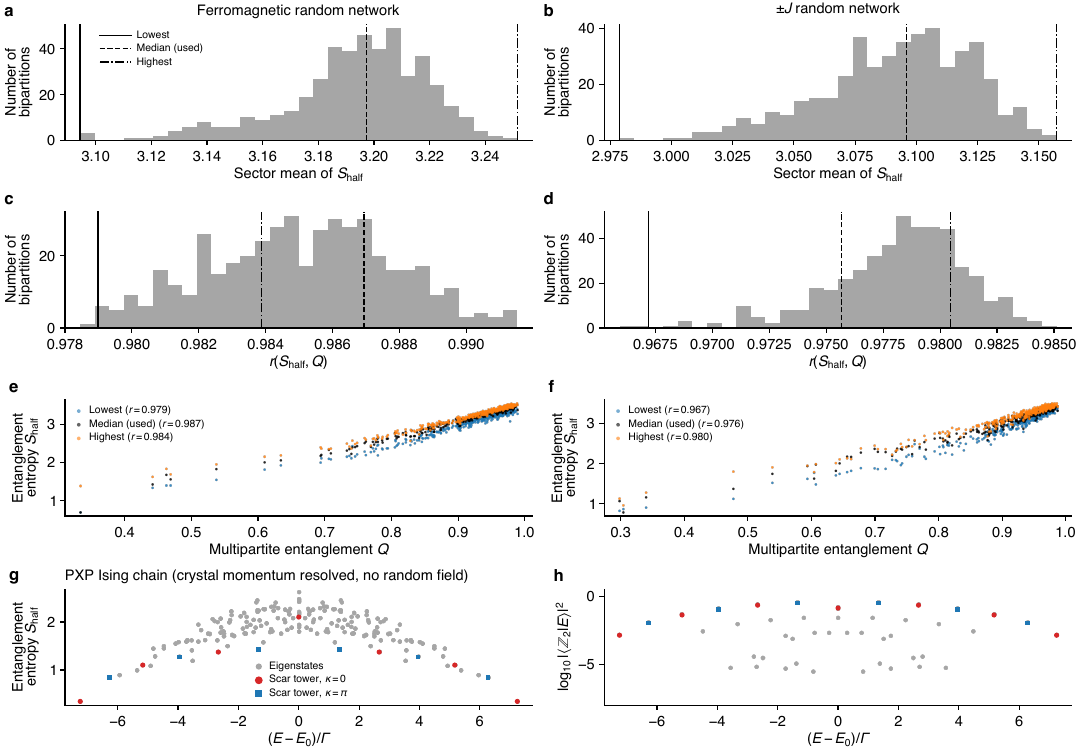}
    \bcaption{Dependence of the half-system entanglement entropy on the
    bipartition, and the scar tower of the ring.}{%
        a--f, Random networks.
        Left column (a,c,e): ferromagnetic random network;
        right column (b,d,f): $\pm J$ random network,
        in the same subspaces as in Extended Data Fig.~\ref{fig:fig-s05}.
        Each network is cut into two sets of $N/2 = 6$ spins in all
        $\binom{N}{N/2}/2 = 462$ ways.
        a,b, Histogram of $S_{\mathrm{half}}$ averaged over the
        eigenstates of the sector.
        c,d, Histogram of the Pearson correlation coefficient $r$ between
        $S_{\mathrm{half}}$ and the multipartite entanglement $Q$ over
        the eigenstates of the sector.
        Solid, dashed and dash-dotted vertical lines mark the
        bipartitions with the lowest, median and highest sector average
        of $S_{\mathrm{half}}$, respectively;
        the median bipartition is the one used in
        Extended Data Fig.~\ref{fig:fig-s05}g,h,m,n.
        e,f, $S_{\mathrm{half}}$ against $Q$ for the three bipartitions
        marked in a--d:
        lowest (blue), median (black) and highest (orange).
        Degenerate eigenstates are excluded as in Extended Data
        Fig.~\ref{fig:fig-s05}.
        g,h, Ring:
        sector of the N\'eel state of the ring ($M = 322$),
        diagonalized in the basis of the crystal momentum $\kappa$.
        g, $S_{\mathrm{half}}$, averaged over the $N/2 = 6$ contiguous
        half-ring cuts,
        against the eigenenergy measured from the classical energy $E_0$
        of the sector in units of $\varGamma$.
        h, Overlap $|\langle\mathbb{Z}_2|E\rangle|^2$ with the two
        N\'eel states;
        states at $\kappa \neq 0, \pi$ have zero overlap and lie below the
        plotted range.
        Red circles ($\kappa = 0$) and blue squares ($\kappa = \pi$) mark the
        $L + 1 = 13$ states of the scar tower~\cite{Turner2018a}:
        starting from the zero-energy state of largest overlap,
        each is the $\kappa = 0$ or $\pi$ state of largest overlap between
        $0.5$ and $1.5$ times the preceding spacing away from the
        previous one.
    }
    \label{fig:fig-s06}
\end{figure*}

\begin{figure*}[tb]
    \centering
    \includegraphics{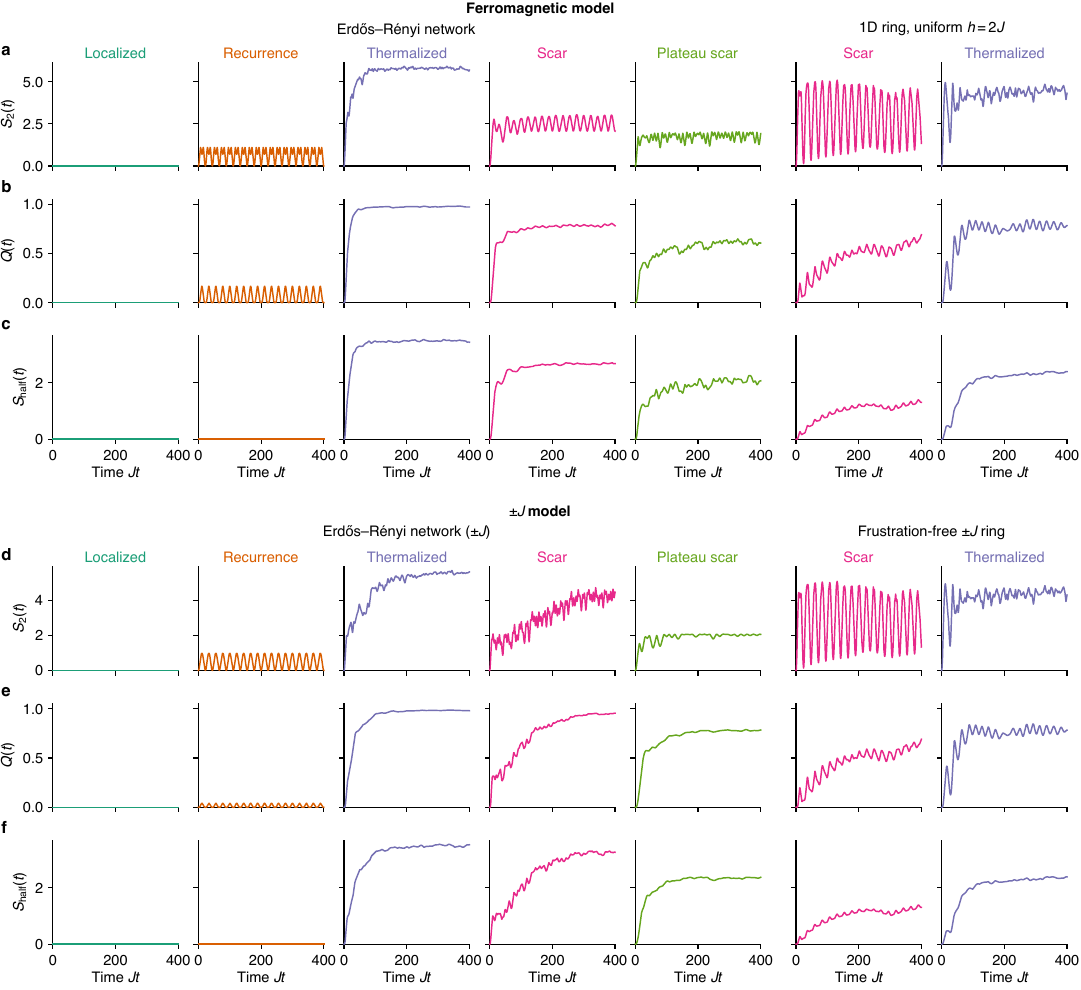}
    \bcaption{
        Time evolution of the participation entropy, the multipartite
        entanglement, and the half-system entanglement entropy.
    }{
        $S_2(t)$, $Q(t)$, and $S_{\mathrm{half}}(t)$ of
        $\ket{\psi(t)} = \exp(-\mathrm{i} \hat{\mathcal{H}} t) \ket{\bm{s}_0}$
        for the initial Fock states of Fig.~\ref{fig:fig-03} of the main
        text (\textbf{a}--\textbf{c})
        and of Extended Data Fig.~\ref{fig:fig-s03}
        (\textbf{d}--\textbf{f}),
        with the same columns and class colours
        ($\varGamma = 0.1J$, $N = 12$).
        Each state evolves within its own sector.
        \textbf{a}, \textbf{d},
        Rényi-2 participation entropy $S_2(t)$.
        \textbf{b}, \textbf{e},
        Multipartite entanglement $Q(t)$.
        \textbf{c}, \textbf{f},
        Half-system von Neumann entanglement entropy
        $S_{\mathrm{half}}(t)$,
        with the bipartitions of Extended Data Fig.~\ref{fig:fig-s05}:
        the balanced bipartition whose sector-averaged
        $S_{\mathrm{half}}$ is the median over all bipartitions for the
        random networks,
        and the average over the contiguous half-ring cuts for the
        rings.
        The $x$ axis is shared within each column,
        and the $y$ axis within each row.
    }
    \label{fig:fig-s07}
\end{figure*}

\medskip
\noindent\textbf{Supplementary Movie 2 \textbar{} Time evolution of the
dynamical classes of Extended Data Fig.~\ref{fig:fig-s03}.}
As Supplementary Movie 1,
for the $\pm J$ network and the Mattis-gauge $\pm J$ ring of
Extended Data Fig.~\ref{fig:fig-s03}.

\medskip
\noindent\textbf{Supplementary Movie 3 \textbar{} Sequential spin flips of
the scar state of Fig.~\ref{fig:fig-03}.}
The scar column of Fig.~\ref{fig:fig-03},
with its Fock-space landscape replaced by the reduced Hilbert space of
spins $4$, $5$, and $12$,
the three spins with the largest oscillation of
$\langle \hat{Z}_i(t) \rangle$.
Node colour is the marginal probability of each of the $2^3$
configurations of these spins,
with the other nine spins traced out.
Nodes are arranged by Hamming distance from the initial configuration.
Duration $60$~s at $30$ frames per second.

\bibliography{references}

\clearpage

\clearpage

\clearpage

\clearpage

\clearpage

\clearpage

\clearpage

